\documentclass[12pt,a4paper,english]{article}

\usepackage[bindingoffset=0.3cm,textheight=22.5cm,hdivide={2.7cm,*,2.7cm}, vdivide={*,22cm,*}]{geometry}
\usepackage[dvips,bookmarksnumbered=true,breaklinks=true]{hyperref}

\usepackage{amsmath,bm,amsfonts,amssymb,babel,slashed}

\usepackage[numbers,square,comma,sort&compress]{natbib}

\IfFileExists{dsfont.sty}
	{\usepackage{dsfont}
         \newcommand{\id}{\mathds{1}}}
	{\typeout{Package dsfont.sty was not found, using alternative macros.}
         \let\mathds=\mathbb
         \newcommand{\id}{\mbox{1 \kern-.59em {\rm l}}}}
\let\one=\id

\makeatletter

         \@addtoreset{equation}{section}
\makeatother
\let\startappendix=\appendix
\newcommand{\nocontentsline}[3]{}
\newcommand{\tocless}[3]{\bgroup\let\addcontentsline=\nocontentsline#1{#2}#3\egroup}

\newcommand{\Appendix}[1]{
  \refstepcounter{section}
  \section*{Appendix \thesection:\hspace*{1.5ex} #1}
  \addcontentsline{toc}{section}{Appendix \thesection}
}
\newcommand{\qed}{\nobreak \ifvmode \relax \else
      \ifdim\lastskip<1.5em \hskip-\lastskip
      \hskip1.5em plus0em minus0.5em \fi \nobreak
      \vrule height0.75em width0.5em depth0.25em\fi}

\newcommand{\be}{\begin{equation}}
\newcommand{\ee}{\end{equation}}
\newcommand{\eq}[1]{(\ref{#1})}

\def\nn{\nonumber}
\def\bea{\begin{eqnarray}}
\def\eea{\end{eqnarray}}
\def\obar{\overline}

\def\beqa{\begin{eqnarray}} 
\def\eeqa{\end{eqnarray}} 
\def\beq{\begin{equation}} 
\def\eeq{\end{equation}}

\def\Tr{{\rm Tr}}

\def\a{\alpha}          
\def\b{\beta}           
\def\d{\delta}

\def\g{\gamma}

\def\l{\lambda} \def\L{\Lambda}

\def\s{\sigma}  

\def\th{\theta}

\def\cA{{\cal A}}  \def\cC{{\cal C}}
  \def\cF{{\cal F}}
\def\cG{{\cal G}} \def\cH{{\cal H}} \def\cI{{\cal I}}
 \def\cK{{\cal K}} \def\cL{{\cal L}}
\def\cM{{\cal M}} \def\cN{{\cal N}} \def\cO{{\cal O}}
\def\cP{{\cal P}}  
 \def\cT{{\cal T}}

\newcommand{\R}{\mathds{R}}

\def\bit{\begin{itemize}}
\def\eit{\end{itemize}}

\def\({\left(}
\def\){\right)}
\def\diag{\mbox{diag}}

\def\d{\delta}

\def\pa{\partial} \def\del{\partial}

\def\bcomment#1{}

\def\LNC{\Lambda_{\rm NC}}
\def\YM{{\rm YM}}

\newcommand{\nc}{non-com\-mu\-ta\-tive}

\newcommand{\co}[2]{[#1,#2]}						% commutator
\renewcommand{\a}{\alpha}
\renewcommand{\b}{\beta}
\renewcommand{\d}{\delta}

\renewcommand{\th}{\theta}

\renewcommand{\l}{\lambda}
\newcommand{\G}{\Gamma}

\renewcommand{\L}{\Lambda}
\renewcommand{\Xi}{\Xi}
 \allowdisplaybreaks[3]

\begin{document}

\renewcommand{\title}[1]{\vspace{10mm}\noindent{\Large{\bf
#1}}\vspace{8mm}} \newcommand{\authors}[1]{\noindent{\large
#1}\vspace{5mm}} \newcommand{\address}[1]{{\itshape #1\vspace{2mm}}}

%\thispagestyle{hepth}

%\begin{titlepage}
\begin{flushright}
TUW-12-05\\
CCNY-HEP-12/4
%arXiv:0706.0398\\
\end{flushright}

\begin{center}

\title{ \Large  Gravity and compactified branes in matrix models}

\vskip 3mm

\authors{Harold {Steinacker{\footnote{harold.steinacker@gmail.com}}}
}

\vskip 3mm

\address{ {\it Institute for Theoretical Physics, Vienna University of Technology\\
 Wiedner Hauptstrasse 8-10, A-1040 Vienna (Austria)  } \\[1ex]
and \\[1ex]
{\it  Physics Department\\
City College of the City University of New York\\
160 Convent Avenue, New York, NY 10031 }
}

\vskip 1.4cm

\textbf{Abstract}

\end{center}

A mechanism for emergent gravity on brane solutions in Yang-Mills matrix models is exhibited. Gravity 
and a partial relation between the Einstein tensor and the energy-momentum tensor 
can arise from the basic matrix model action, without invoking an Einstein-Hilbert-type term. 
The key requirements are compactified extra dimensions with extrinsic curvature 
$\cM^4 \times \cK \subset \R^D$ and split 
noncommutativity, with a Poisson tensor $\theta^{ab}$ linking the compact with the noncompact directions. 
The  moduli of the compactification provide the dominant degrees of freedom for gravity, 
which are transmitted to the 4 noncompact directions via the Poisson tensor. 
The effective Newton constant is determined by the scale of noncommutativity and the compactification.
This gravity theory is well suited for quantization, and argued to be perturbatively finite
for the IKKT model. Since no compactification of the target space is needed,
it might provide a way to avoid the landscape problem in string theory.

\vskip 1.4cm

\newpage

\tableofcontents

\section{Introduction}
%%%%%%%%%%%%%%%%%%%%%%%%%%%%%

Matrix models such as the IKKT respectively IIB model \cite{Ishibashi:1996xs}
provide fascinating candidates for a quantum theory of fundamental interactions. 
Part of the appeal stems from the fact that geometry is not an input, but emerges on the solutions. 
For example, it is easy to see that flat noncommutative (NC) planes $\R^{2n}_\theta$ arise as solutions.
Similarly, branes with non-trivial geometry arise as  NC 
sub-manifolds $\cM \subset \R^{10}$, which can be interpreted as physical space-time.  
Their effective geometry is easily understood 
in the semi-classical limit \cite{Steinacker:2008ri,Steinacker:2010rh}, 
in terms of a dynamical effective metric $G_{ab}$ which is strongly reminiscent of the
open string metric in the presence of a $B$- field \cite{Seiberg:1999vs}. 
This metric governs the kinematics of all propagating fields on the brane, and therefore describes 
gravity on the brane.
Moreover, a relation with IIB supergravity or superstring theory\footnote{for related work on the 
BFSS model \cite{Banks:1996vh} see e.g. \cite{Kabat:1997sa,Chepelev:1998sm,Taylor:1997dy,Bak:2002wy}.} has been conjectured and verified 
to a certain extent \cite{Ishibashi:1996xs,Chepelev:1997av,Mandal:1997kj,Kimura:1999qh,Kitazawa:1998dd,Blaschke:2011qu}.
On the other hand, the IKKT model 
can equivalently be viewed as $\cN=4$ supersymmetric Yang-Mills (SYM) theory 
on $\R^4_\theta$, and is thus (expected to be) perturbatively finite in 4 dimensions.
Combining these two points of view strongly suggests that the model should provide a 
quantum theory of fundamental interactions including gravity in 4 dimensions\footnote{The relation with string theory 
is not the main topic of this paper, and we focus on the 4-dimensional brane geometry. However we will
argue that the bulk should be understood in a holographic manner.}.

However, a single 4-dimensional brane $\cM^4 \subset \R^{10}$ is clearly too simple to reproduce the rich 
spectrum of phenomena in nature. In order to recover e.g. the standard model, additional structure is needed.
One possible origin of such additional structure are compactified extra dimensions, as considered 
in string theory. By considering intersecting branes and compactified extra dimensions in the matrix model, it is indeed
possible to obtain chiral fermions and recover the basic structure of the standard model
 \cite{Chatzistavrakidis:2011gs}, adapting ideas from string theory \cite{Blumenhagen:2006ci}.

The main result of the present paper is to show that
the geometrical degrees of freedom provided by compactified extra dimensions 
also play a key role for the effective (emergent) gravity on such branes. 
Note that although the geometry of the branes is easy to describe, the dynamics of this geometry is complicated and not
well understood. The central point is that the basic degrees of freedom are different from GR: 
the effective metric $G^{ab}$ on the brane is not fundamental, but determined in terms of the 
brane embedding $\cM \subset \R^{10}$ along with the  Poisson structure 
$\theta^{ab}$ resp. the $B$-field. These basic degrees of freedom are governed by equations of motion of
the matrix model, and it is not evident that general relativity (GR) will emerge on the brane.
A priori the matrix model does not contain an Einstein-Hilbert term, 
although it will be induced by quantum effects. 
Thus an induced gravity mechanism is conceivable, though delicate.
On the other hand, the different degrees of freedom imply that 
even in the presence of an induced Einstein-Hilbert term there will be additional solutions,
which are deformations of the basic
``harmonic'' solutions  \cite{Steinacker:2010rh} of the matrix model. 
%It is thus not evident what type of solutions -- Einstein or harmonic -- are appropriate.
Disentangling these degrees of freedom and understanding their significance is far from trivial\footnote{A different approach to 
obtain gravity from the IKKT model has been proposed in \cite{Hanada:2005vr}. Its significance and relation with the solutions 
considered here is not clear to the author.}.

In the present paper, we exhibit a novel mechanism for gravity on branes with 
compactified extra dimensions $\cM^4 \times \cK \subset \R^{10}$, 
which is based on the bare matrix model action without 
requiring the presence of an induced Einstein-Hilbert type term. The mechanism 
is therefore  robust, and is expected to give the dominant contribution to 4-dimensional 
gravity on the brane. Although the brane gravity is {\em not} equivalent to GR,
essential features of GR are recovered, in particular a partial relation between the 
Einstein tensor and the energy-momentum (e-m) tensor. The precise coupling  depends on the compactification,
and we focus mainly on the simplest case of $\cM^4 \times T^2$ in the present paper. 
Newtonian gravity  is recovered, with an effective Newton constant 
determined by the scale of noncommutativity and the compactification. 
This mechanism discussed depends on two crucial conditions:
\begin{enumerate}
 \item 
non-vanishing extrinsic curvature  of the embedding $\cM = \cM^4 \times \cK \subset \R^{10}$
predominantly due to $\cK \subset \R^{10}$, and
\item
''split noncommutativity`` \cite{Steinacker:2011wb} where the Poisson tensor $\theta^{ac}$ 
links the noncompact with the compact spaces. This 
transmutes perturbations of the moduli of $\cK$ into perturbations of the effective metric.
\end{enumerate}
Let us discuss this in  more detail. In GR,
gravity is characterized the intrinsic geometry of the 4-dimensional space-time manifold, while its specific 
realization -- via an isometric embedding or as abstract  manifold -- is irrelevant. 
For branes arising as solutions of  matrix models, this is not the case.
% First and foremost, the fundamental degrees of freedom are different, given not by the metric but by 
% brane embeddings and their fluctuations, as well as a Poisson structure.
% The metric is  a derived quantity.
One key observation  \cite{Steinacker:2009mp}
is that linearized embedding fluctuations couple 
linearly to the energy-momentum tensor in the presence of 
{\em extrinsic curvature} of $\cM \subset \R^{10}$, leading to (Newtonian, at least) gravity.
However this required somewhat ad-hoc assumptions. 
The new observation in the present paper is that the mechanism naturally applies for
compactified extra dimensions, leading to a (partial) relation between the Einstein tensor and the energy-momentum tensor
at least in the linearized approximation. 
We also argue that Ricci-flat vacuum geometries arise naturally albeit not necessarily.
Moreover, fluctuations of the compactification moduli of
$\cK$ are transmuted via  $\theta^{ab}$ to 4-dimensional metric fluctuations.
Here $\cK$ is in a sense rotating and stabilized by angular momentum, 
given e.g. by a torus $\cK=T^n$ with light-like compactification.
Such solutions of the matrix model have been given recently \cite{Steinacker:2011wb}.

It is important to emphasize that the effective gravity is indeed 4-dimensional, even though the brane
$\cM^4 \times \cK \subset \R^{D}$ is embedded in a higher-dimensional non-compact target space.
This is in contrast to the conventional picture in string theory, where gravity is supposed to originate from 
closed strings which  propagate in 10 dimensions, leading to a 10-dimensional Newton law. 
The crucial point is that brane gravity emerges here\footnote{This was also realized recently in a different approach to 
emergent gravity \cite{Heckman:2011qu}, however their matrix model still requires UV completion via string theory.
The present model is claimed to be complete. The origin for the 4-dimensional behavior is also very different from e.g. 
the DGP mechanism \cite{Dvali:2000hr}, which is based on a combination of brane and bulk physics without a $B$ field.}  
entirely within the open string sector on the brane with a non-vanishing Poisson structure resp. $B\neq 0$, 
while the 10-dimensional bulk 
arises in a holographic manner (cf. \cite{Maldacena:1997re}; this can  be seen in the matrix model
at the one-loop level \cite{Chepelev:1997av,Blaschke:2011qu}). 
This is very welcome, since one can now discard the vast landscape of 10-dimensional compactifications,
and study the mini-landscape of embedded compactified branes embedded in $\R^{10}$, as described by the 
IKKT model. This is a well-posed problem which should have a clear non-perturbative answer.

At first glance, a non-trivial Poisson structure $\theta^{ab}$ resp. $B$-field on the brane 
may seem incompatible with Lorentz invariance. However, 
 $\theta^{ab}$ is completely absorbed in the effective metric $G^{ab}$ and does not explicitly
couple to any field at least at tree level \cite{Steinacker:2010rh}. 
Therefore the effective action for all propagating fields on the brane is compatible with (local)
Lorentz transformation as defined by $G^{ab}$. On the other hand, the dynamics of the 
geometry is sensitive to $\theta^{ab}$, which may lead to a gravitational 
violation of (local) Lorentz invariance and a violation of the equivalence principle.
Indeed anisotropic post-Newtonian corrections seem to arise in the simplified analysis given here.
We argue that more sophisticated backgrounds and/or a more complete treatment should alleviate this problem, so that 
a viable gravity could be obtained from this mechanism.
As a related bonus, we will argue that the physics of vacuum energy is different from GR, which
could have important consequences for cosmology. We  give an argument that the usual fine-tuning 
problem associated with quantum mechanical vacuum fluctuations
leading to the cosmological constant problem does not arise here.

Finally, although the backgrounds $\cM^4 \times \cK$ under consideration 
are at least 6-dimensional at low energies, they  behave as 4-dimensional spaces in the UV due to 
noncommutativity \cite{Steinacker:2011wb}. Then the IKKT model can be viewed as
$\cN=4$ SYM in 4 dimensions, which is expected to be UV finite and without pathological UV/IR mixing \cite{Jack:2001cr}. 
Therefore the 
the model is expected to be UV finite for the backgrounds under consideration. This is in stark contrast with 
commutative SYM and supergravity, which would not be UV complete and require a UV
completion via string theory. The matrix model approach therefore promises to provide 
a consistent and self-contained approach towards a quantum theory of all fundamental interactions,
and might resolve the landscape problem in string theory.
This  certainly justifies  further studies.

This paper is organized as follows. After a brief review of basic aspects of branes in matrix models,
we discuss in detail  perturbations of branes and the associated curvature perturbations. 
The dynamics of these metric fluctuations coupled to matter is derived in section 4.
Next the compactification is discussed in some detail, and the role of moduli 
as mediators of gravity is exhibited.
The effective 4-dimensional gravity is then derived, focusing on the case of toroidal compactifications.
In particular the background $M^4 \times T^2$ is worked out in detail. 
In section 5 we discuss possible generalizations of the compactification, arguing that 
more realistic backgrounds can be found. However, the study of such generalized compactifications 
is left for future work.

Throughout this paper, a slightly cumbersome but explicit index notation is used. This is done 
in order not to hide things under the carpet, and a more elegant formulation can be given eventually.

\section{Matrix models and their geometry} 
\label{sec:matrixmodels-intro}
%%%%%%%%%%%%%%%%%%%%%%%%%%%%%%%%

We briefly collect the essential ingredients of the matrix model framework
and its effective geometry, referring 
to the recent review \cite{Steinacker:2010rh} for more details.

\subsection{The IKKT model and related matrix models}
\label{sec:basic}

The starting point is given by a matrix model of Yang-Mills type, 
\begin{align}
S_{\YM} &=-\frac {\L_0^4}4\Tr\(\co{X^A}{X^B}\co{X^C}{X^D}\eta_{AC}\eta_{BD}\,  +  2 \obar\Psi \gamma_A[X^A,\Psi] \)   
\label{S-YM}
\end{align}
where the $X^A$ are Hermitian matrices, i.e. operators acting on a separable Hilbert space $\mathcal{H}$. The
indices of the matrices run from $0$ to $D-1$, and
will be raised or lowered with  the invariant tensor $\eta_{AB}$ of $SO(D-1,1)$.
Although this paper is mostly concerned with the bosonic sector, we focus 
 on the maximally supersymmetric IKKT or IIB model \cite{Ishibashi:1996xs} with $D=10$, which is best suited for quantization.
Then $\Psi$ is a matrix-valued Majorana Weyl spinor of $SO(9,1)$.
The model enjoys the fundamental gauge symmetry 
\be
X^A \to U^{-1} X^A U\,,\qquad  \Psi \to U^{-1} \Psi U\,,\qquad    U \in U(\cH)\,
\label{gauge-inv}
\ee
as well as the 10-dimensional Poincar\'e symmetry
\be
\begin{array}{lllll}
X^A \to \L(g)^A_B X^b\,,\quad  &\Psi_\a \to \tilde \pi(g)_\a^\b \Psi_\b\,,\quad  & g \in \widetilde{SO}(9,1), &  \\[1ex]
X^A \to X^A + c^A \one\,,\quad & \quad & c^A \in \R^{10}\,\quad % & \mbox{translational symmetry,}
\end{array}
\label{poincare-inv}
\ee
and a $\cN=2$ matrix supersymmetry \cite{Ishibashi:1996xs}.
The tilde indicates the corresponding spin group.
We also introduced a parameter  $\L_0$ of dimension $[L]^{-1}$, so that the  $X^A$ have dimension length,
corresponding to the (trivial) scaling symmetry
\begin{align}
X^A \to  \a X^A, \qquad \Psi \to  \a^{3/2}\Psi,  \qquad \L_0 \to \a^{-1}\L_0 .
\label{scaling}
\end{align}
On backgrounds with $S=0$ there is also a non-trivial scaling  symmetry 
$X^A\to \a X^A$.
We  define the matrix Laplacian as
\begin{align}
  \Box \Phi :=  [X_B,[X^B,\Phi]]
\label{box-def}
\end{align}
for any matrix $\Phi \in \cL(\cH)$.
Then  the equations of motion of the model take the following form
\be
 \Box X^A \,=\, [X_B,[X^B,X^A]] =  0
\label{eom-IKKT}
\ee
for all $A$, assuming  $\Psi = 0$.

\subsection{Noncommutative branes and their geometry}

Now we focus on matrix configurations which describe embedded
noncommutative (NC) branes. This means that 
the $X^A$ can be interpreted as quantized embedding functions \cite{Steinacker:2010rh}
\be
X^A \sim x^A: \quad \mathcal{M}^{2n}\hookrightarrow \R^{10} 
\ee 
of a $2n$- dimensional submanifold of $\R^{10}$. More precisely,
there should be some quantization map $\cI: \cC(\cM)  \to  \cA\subset L(\cH)$
which maps classical functions on $\cM$ to a noncommutative (matrix) algebra of functions, 
such that commutators can be interpreted as quantized Poisson brackets. In the
semi-classical limit indicated by $\sim$,  matrices are identified with functions via $\cI$,
and commutators are replaced by Poisson brackets; for a more extensive introduction
see e.g. \cite{Steinacker:2010rh,Steinacker:2011ix}.
One can then locally choose $2n$ independent coordinate functions
$x^a,\ a = 1,...,2n$ among the $x^A$, and their commutators
\begin{align}
\co{X^a}{X^b} \ &\sim \  i \{x^a,x^b\} \ = \ i \th^{ab}(x)\,
\end{align}
encode a quantized Poisson structure on $(\cM^{2n},\theta^{ab})$.
These $\theta^{ab}$ have dimension $[L^2]$ and set a typical scale of noncommutativity $\LNC^{-2}$.
We will assume that $\th^{ab}$ is non-degenerate\footnote{If the Poisson structure is degenerate, then fluctuations propagate only
along the symplectic leaves.}, so that the 
inverse matrix $\th^{-1}_{ab}$ defines a symplectic form 
on $\mathcal{M}^{2n}\subset\R^{10}$. This submanifold 
is equipped with the induced metric
\begin{align}
g_{ab}(x)=\pa_a x^A \pa_b x_A\,
\label{eq:def-induced-metric}
\end{align}
which is the pull-back of $\eta_{AB}$.  However, this is {\em not} the effective metric on $\cM$. 
To understand the effective metric and gravity, we need to consider matter on the brane $\cM$. 
Bosonic matter or fields arise from 
nonabelian fluctuations of the matrices around a stack $X^A \otimes \one_n$
of coinciding branes, while fermionic matter arises
from $\Psi$ in \eq{S-YM}.
It turns out that  in the semi-classical limit, the  effective action for such fields is
governed by a universal effective metric $G^{ab}$. 
It can be obtained most easily by considering the action of an additional %(for the sake of demonstration) 
scalar field $\phi$ coupled to the matrix model in a gauge-invariant way, with action
\begin{align}
 S[\phi] &= -\frac{\L_0^4}2\Tr\, [X_A,\phi][X^A,\phi] 
\sim \frac{\L_0^4}{2(2\pi)^n} \int d^{2n} x \sqrt{|\theta^{-1}|} \th^{aa'}\th^{bb'}g_{a'b'}\, \del_a\phi \del_b \phi \nn\\
 &= \frac{\L_0^4}{2(2\pi)^n}  \int d^{2n} x\sqrt{| G|} \, G^{ab} \del_a\phi \del_b \phi .
 \label{action-scalar-geom}
\end{align}
Therefore  the effective metric is given by \cite{Steinacker:2008ri}
\begin{align}
 G^{ab}&= e^{-\sigma} \th^{aa'}\th^{bb'}g_{a'b'}\,,   \nn\\
  e^{-\s}&=\Big(\frac{\det{\th^{-1}_{ab}}}{\det{ G_{ab}}}\Big)^{\frac 1{2}}\,  
 = \Big(\frac{\det{\th^{-1}_{ab}}}{\det{g_{ab}}}\Big)^{\frac 1{2(n-1)}}   % =: \LNC^{\a} .
\label{eff-metric}
\end{align}
which is very much like the open string metric on D-branes with a $B$-field \cite{Seiberg:1999vs}.
Let us briefly discuss the scales and dimensions. Clearly $e^{-\s}$ characterizes the 
NC scale\footnote{Thus $e^{-\s}$ is dimensionful, and could be made
dimensionless by absorbing suitable powers of $\L_0$. However we stick to the above conventions to
keep the formulae simple.
Note also that the scaling dimensions of $G$ and $g$ are distinct and rather peculiar for $\dim\cM \neq 4$.
This reflects the fact that the trace is related to the symplectic volume form.}, 
and $\L_0$ is related to $e^{-\s}$ via the  Yang-Mills coupling constant \cite{Steinacker:2010rh}
\begin{align}
 \L_0^{4} e^\s = \frac 1{g_{\YM}^2} 
\label{gauge-coupling}
\end{align}
which governs the $SU(n)$ sector. 
The transversal matrices resp. their fluctuations $\phi^i\equiv \d X^i$ will be considered as perturbation of the embedding, 
with dimension $\dim\phi^i = [L]$. On the other hand,  nonabelian fluctuations of the transversal 
matrices should be viewed as scalar fields, via the identification 
\begin{align}
 \varphi = \L_0^2 \phi,\qquad  
 S[\phi] \sim \frac{1}{2(2\pi)^n}  \int d^{2n} x \sqrt{| G|} \, G^{ab} \del_a\varphi \del_b \varphi .
\label{action-scalar-field-phys}
\end{align}
Then  $\dim\varphi^i = [L^{-1}]$, and the
energy-momentum (e-m) tensor is recovered correctly. 

The important point is that the metric $G$
governs the semi-classical limit of all fields propagating on $\cM$ including
scalar fields, non-Abelian gauge fields and fermions  \cite{Steinacker:2008ri,Steinacker:2008ya}.
%The latter must be constant to a very good approximation, in view of \eq{gauge-coupling}.
This means that $G$ must be interpreted as gravitational metric. Therefore the model provides a dynamical
gravity theory, realized on dynamically 
determined branes $\cM \subset \R^{10}$  governed by the action \eq{S-YM}. 
To understand the dynamics of the geometry in more detail, 
the following result is  useful \cite{Steinacker:2010rh}: the matrix Laplace operator 
reduces  in the semi-classical limit to the covariant Laplace operator 
\begin{align}
 {\bf\Box} \Phi &= [X_A,[X^A,\Phi]]  
\ \sim\  -e^\s \Box_{G}\, \phi 
\end{align}
acting on scalar fields $\Phi \sim \phi$.
In particular, the matrix equations of motion \eq{eom-IKKT} take the  simple form
\be
0 =  \Box X^A \,\sim\, -e^\s \Box_{G} x^A .
\label{eom-IKKT-semi}
\ee
This means that the embedding functions $x^A \sim X^A$ are harmonic functions with respect to $G$.
Furthermore, the bosonic matrix model action \eq{S-YM} can be written in the semi-classical limit as follows
\begin{align}
 S_{\YM}\ \sim \ \frac{\L_0^4}{4(2\pi)^{2n}}\int d^{2n} x \sqrt{|\theta^{-1}|}\, \g^{ab}g_{ab} .
\label{action-semiclass}
\end{align}
Here we introduce
the  conformally equivalent  metric\footnote{More abstractly, this can be stated as 
$(\a,\b)_\g = (i_\a\theta,i_\b \theta)_g$ where $\theta = \frac 12\theta^{ab}\del_a\wedge\del_b$.}
\begin{align}
\g^{ab}&= \th^{aa'}\th^{bb'}g_{a'b'}\, \ = \ e^\s \, G^{ab}
\label{eff-metric-simple}
\end{align}
which satisfies
\begin{align}
\sqrt{|\theta^{-1}|} \g^{ab} &= \sqrt{|G|} \,G^{ab}  .
\end{align}
Note also that $\theta^{ab}\del_b$ is somewhat reminiscent of a 
frame, which could be 
made more suggestive in the form 
\begin{align}
e^{(a)} = \{x^a,.\} = \theta^{ab} \del_b , \qquad \g^{ab}  = (e^{(a)},e^{(b)})_g  
\label{frame}
\end{align}
cf. \cite{Yang:2010kj}.
However the analogy is somewhat misleading for proper submanifolds,
because $g_{ab}$ is dynamical and not flat in general.

\subsubsection{Perturbations of the matrix geometry}
\label{sec:perturbations}

Now consider a brane $\cM^{2n}\subset \R^D$ obtained as a perturbation  
\begin{align}
X^A = \bar X^A + \d X^A 
\end{align}
of some background brane $\bar \cM$, defined in terms of matrices $\bar X^A$ as above.
We want to understand the  metrics $G^{ab}$ and $g_{ab}$ on $\cM$ as  deformations of 
$\bar G^{ab}$ and $\bar g_{ab}$ on $\bar \cM$. In the semi-classical limit,
the perturbation can be split into $D-2n$ transversal perturbations $\d_\perp X^A \sim  \phi^A$ and 
$2n$ tangential perturbation 
$\d_\|  X^A \sim \cA^A$, defined by 
\begin{align}
\phi_A \del_a \bar x^A &= 0,  \label{transversal-fluct}\\
\cA^A &=  \cA^{a} \del_a \bar x^A . \label{parallel-fluct}
\end{align}
To make this more transparent, consider some point $p\in\bar\cM$.
We can assume using Poincar\'e invariance that $p$ is at the origin, and 
the tangent plane $T_p \bar \cM$ is embedded along the first $2n$ Cartesian embedding coordinates
\be
\bar x^A= (\bar x^a,\bar y^i) \quad \mbox{with}\quad y^i|_p = 0, \ \   \del_a|_p \bar y^i = 0 .
\label{NEC}
\ee
This defines ``normal embedding coordinates'' (NEC) 
$\bar x^a \sim \bar X^a$,
which can be used both on $\bar \cM$ and $p \in \cM$ near  $p$; hence we  omit the bar 
from now on. They are  normal coordinates corresponding to the connection $\nabla^{(g)}$ defined by the embedding metric $g$. 
However, we will mostly use the connection $\nabla\equiv \nabla^{(G)}$
defined by the effective metric $\bar G$ for the following.
Then the transversal variations are given by the $\phi^A = (0,\phi^i)$, and the tangential variations 
by $\cA^A = (\cA^a,0)$.
If the  matrix model is viewed as NC gauge theory, then these variations $\cA^a$ and  $\phi^i$ 
can be interpreted in terms of 
``would-be'' $U(1)$ gauge fields and scalar fields on $\bar\cM$; this is useful for perturbative computations.
The tangential variations lead to a perturbed  Poisson structure on $\cM$ 
\begin{align}
\theta^{ab}(x)  &= \bar\theta^{ab} + \d \theta^{ab}(x) , \qquad
\d \theta^{ab}(x) = \bar\theta^{ac}\bar \theta^{bd} \d\theta^{-1}_{cd} ,
\label{poisson-fluctuation-1}
\end{align}
which can be parametrized in terms of the a ``would-be'' $U(1)$ gauge field 
\be
\d \theta^{-1}_{ab} = F_{ab} =
\nabla_a \d A_b - \nabla_b \d A_a .
\ee
The transversal embedding perturbation
$\d_\perp X^A \sim \phi^A$ satisfy the constraint \eq{transversal-fluct}, which implies
\begin{align}
\del_b \phi_A \del_{a} \bar x^A &= - \phi_A \nabla_b \del_{a} \bar x^A = - \phi_A  K^A_{ab} .
\label{transversal-constraint}
\end{align}
Here
\be
 K^A_{ab} = \nabla_a\del_b \bar x^A = K_{ba}^A
\ee
is the the 2nd fundamental form,
which characterizes the exterior curvature of $\cM \subset \R^D$, and will play a central role in the following.
Notice that $K_{ab}^A$ takes values in the normal bundle $N \bar\cM$ provided $\nabla_c \bar \theta^{ab} = 0$
(since then the connections defined 
by the induced and the effective metrics on $\bar \cM$ coincide, so that
$\del_a x_A\nabla_{b}\del_{c} x^A= 0$).
The metric fluctuations are obtained as
\begin{align}
g_{ab} \ &= \ \bar g_{ab} \ + \ \d g_{ab} , \nn\\
 \d g_{ab}\  &= \ \del_{a} \phi_A \del_{b}\bar x^A +  \del_{a} \bar x_A \del_{b}\phi^A 
 = - 2\phi_A K_{ab}^A ,
\label{del-g-explicit}\\
 \g^{ab} \ &= \ \bar \g^{ab} \ + \ \d \g^{ab},  \nn\\
\d \g^{ab} \ &= \ \bar\theta^{ac} \bar\theta^{bd} \d g_{cd} - \bar\theta^{ac} F_{cc'}  \g^{bc'}  - \bar\theta^{bc} F_{cc'}  \g^{ac'} ,
\label{del-G-explicit}\\
\d \s\  &= \ \frac 1{2}(G^{ab}\d G_{ab} + \bar\theta^{ab} F_{ab})
 =\frac 1{2(n-1)}(g^{ab}\d g_{ab} + \bar\theta^{ab} F_{ab})  ,
\label{del-sigma}
\end{align}
using \eq{eff-metric} in the last equation. There are also quadratic terms in $\phi,F$ which are omitted.
Hence the perturbation of the effective metric $G^{ab}$ is given by
\begin{align}
  \d  G^{ab} &=   e^{-\s} \d \g^{ab} - G^{ab} \d \s   \nn\\
 &= -2 e^{-\s} \bar\theta^{aa'} \bar\theta^{bb'} \Pi^{cd}_{ab} K_{cd}^A\,\phi_A
-  \bar\theta^{ac} F_{cd}  G^{bd}  -  \bar\theta^{bc} F_{cd}  G^{ad} 
 - \frac{ G^{ab}}{2(n-1)} (\bar\theta F)
\label{dtildeG-form}
\end{align}
using the abbreviations
\begin{align}
(\theta F) &= \theta^{ab}F_{ab}  \nn\\
 \Pi^{cd}_{ab} &= \d^{cd}_{ab} - \frac{g_{ab}g^{cd}}{2(n-1)} .
\label{pi-def}
\end{align}

\subsection{Curvature perturbations}

Now we take advantage of the special coordinates $x^a \sim X^a, \ a=1,...,2n$ 
provided by some suitable subset of the  matrices $X^A$, for example the NEC defined above.
The equations of motion  $\Box X^a = 0$ in vacuum
implies that these coordinates satisfy the 
harmonic gauge condition, 
\begin{align}
0 &\sim \Box_{ G}\, x^a  = -\Gamma^a
= | G_{cd}|^{-1/2}\partial_b (\sqrt{| G_{cd}|}\,  G^{ba}) .
\end{align}
For the metric fluctuations $h_{ab} = \d G_{ab}$ this implies  the harmonic gauge condition
\be
\partial^b  h_{ab} - \frac 12 \partial_a  h =0 .
\label{h-harmonic}
\ee
Then the perturbation of the Ricci tensor 
around a general background  \cite{Wald:1984rg}
\begin{align}
 \d R_{ab} &= -\frac 12 \bar\nabla_a\del_b h - \frac 12 \bar\Box_G h_{ab} + \bar\nabla_{(a}\bar\nabla^d h_{b)d} 
\label{lin-Ricci-general}
\end{align}
simplifies as
\be
\d_h R_{ab}  = - \frac12 \bar\Box_{G}  h_{ab} \ . 
\label{Ricci-harmonic}
\ee
Similarly, the perturbations of the Einstein tensor $\cG_{ab}  =  R_{ab} - \frac 12 G_{ab} R $   can be  written
as follows
\begin{align}
\d\cG_{ab}  &= - \frac12 \bar\Box_{G} \Big(\d G_{ab} - \frac 12 \bar G_{ab} (G^{cd}\d G_{cd})\Big)  \ .
\end{align}
Noting that $\d  G_{ab} = -\bar G_{aa'}\bar G_{bb'} \d  G^{a'b'}$ while $\d\cG_{ab}$ is a tensor, this can be written as 
\begin{align}
 \d\cG^{ab}  &=  \frac12 \Big( \bar\Box_{G} \d G^{ab}  +\frac 12  \bar G^{ab} \bar\Box_{G}(\bar G^{cd}\d G_{cd})\Big) \ \nn\\
 &= \frac12 \Big( \bar\Box_{G} (e^{-\s}\d \g^{ab} - \bar G^{ab} \d \s)  + \bar G^{ab} \bar\Box_{G}(\d\s - \frac 12 (\bar\theta F) \Big)\  \nn\\
 &=  \frac12\Big(e^{-\s}  \bar\Box_{G} \d \g^{ab}  - \frac 12 \bar G^{ab}  \bar\Box_{G} (\theta F) \Big)\ 
\label{einstein-tensor-fluct}
\end{align}
using \eq{del-sigma}, assuming  $\s = const$ for the background. 
We will mostly using this equation for a locally adapted flat background, corresponding to 
normal coordinates. Then the curvature corrections drop out, and 
this relation allows to compute the full Einstein tensor.

\section{Gravity on higher-dimensional branes}

To  understand how matter affects the  geometry, we now
study the dynamics of these geometrical modes. 
The  goal is to show that the vacuum geometry is Ricci-flat to a good approximation,
and matter couples to the Einstein tensor in a way similar to general relativity.

Assume $\cM \subset \R^D$ is a  brane as above, described by $X^A \sim x^A$.
Unlike in general relativity, 
the dynamics is governed by the effective action \eq{action-semiclass}, where 
the geometric perturbations can be organized into 
transversal  and parallel fluctuations $\d_\perp x^A = \phi^A$ resp. $\d \theta^{-1}_{ab} = F_{ab}$ 
as above.
We  expand the action  \eq{action-semiclass} in $\phi^A$ and $F_{ab}$. To first order, one obtains
\begin{align}
 S^{(1)}[\phi,F]  &=  \frac{\L_0^4}{4(2\pi)^n}\! \int\! d^{2n} x \sqrt{|G|} \Big(
 2 F_{ab} \big(e^{\sigma}G^{bc}\theta^{-1}_{cd}G^{d a}\,- \frac 14\theta^{ab}\,(G\cdot g) \big)
 - 4 \phi_A  K_{ab}^A G^{ab} +  \l^a \phi_A \del_a \bar x^A \Big) 
\label{action-expanded-1}
\end{align}
where $(G\cdot g)  \equiv G^{cd}g_{cd}$.
Now we take matter into account.
Physical fields and matter arises in the matrix model from nonabelian fluctuations
of the bosonic matrices $X^A$ around the background, and from the fermionic matrices $\Psi$.
Since they couple in the standard way to the effective metric $G$,  the variation of their action with respect to 
fluctuations of the geometry can be written as\footnote{Notice that there is no scale factor $\L_0$;
it is absorbed in $T_{ab}$ upon recasting the nonabelian fields into physical form, 
introducing dimensions as in \eq{eff-metric}. The normalization is chosen to avoid factors $2\pi$ later.}
\begin{align}
\d S_{\rm matter} &= \frac{1}{2(2\pi)^n} \int d^{2n}x\, \sqrt{ G} \, T_{ab} \,  \d  G^{ab} 
\label{dS-matter}
\end{align}
where $\d  G^{ab}$ is given by \eq{dtildeG-form}.
To obtain the equations of motion for the geometry, we need 
the variation of the matter action 
\begin{align}
\int d^{2n} x \sqrt{G} T_{ab} \, \d_\phi  G^{ab} &= 
-\int d^{2n}x\, \sqrt{ G} \, \phi_A\, \big(e^{-\s} T_{ab} \theta^{aa'} \theta^{bb'} \Pi^{cd}_{a'b'} K_{cd}^A\big) \nn\\
\int d^{2n} x \sqrt{G} T_{ab} \, \d_\cA G^{ab} 
&= \int d^{2n}x\, \sqrt{ G} \,\d A_d\, \big( 2 T_{ab} \, \nabla^b\theta^{da}
 +e^{-\s}\theta^{ac} \nabla_c (e^\s T_{ab})  G^{db} + \frac{e^{-\s}}{n-1} \theta^{cd} \nabla_c T \big)  \nn
\end{align}
where $T =  \g^{ab} T_{ab}$, 
dropping a term proportional to\footnote{The usual form of the conservation law is assumed, although
it might be slightly modified in the NC case.} $\nabla^a T_{ab}$, and
using the identity \cite{Steinacker:2010rh}
\begin{align}
\nabla_a (e^{-\s} \theta^{ab}) = 0 .
\label{theta-id}
\end{align}
Instead of writing an equation for $\theta^{ab}$ in the form 
$\nabla^d(e^{\sigma}\theta^{-1}_{cd}) - \frac 14 G_{bc}\theta^{ab}\,\del_a (G\cdot g) \,= \cO(T)$ as in
 \cite{Steinacker:2008ri}, it is more useful  to rewrite $S^{(1)}[\phi,F]$ 
using the identity \eq{geom-id-cons} as
\begin{align}
S^{(1)}[F] &= \frac{\L_0^4}{(2\pi)^n} \int  d^{2n}x\, \,\Big(
 A_a \nabla_b (\sqrt{| G|}\, G^{bc} T^{\rm geom}_{cd} \theta^{da}) \Big) \nn\\
 &= -\frac{\L_0^4}{(2\pi)^n} \int  d^{2n}x\,\sqrt{G} \,A_a\Big( e^{\s}  G^{ab} G^{de}\nabla_e^{(g)}\theta^{-1}_{db} \Big)
\end{align}
where $T^{\rm geom}$ is defined in \eq{T-geom-def}. 
We then obtain the equations of motion 
\begin{align}
G^{ab}\nabla_a^{(g)}\nabla_b^{(g)} x^A \ &= \ - \L_0^{-4}\, e^{-\s}  K_{cd}^A  \Pi^{cd}_{a'b'}\theta^{aa'} \theta^{bb'} \  T_{ab}   , \label{eom-K-T}\\
% = \Box_G x^A   + \l^a \del_a  x^A  
G^{ce}\nabla_e^{(g)}\theta^{-1}_{cb} \ &= \ \L_0^{-4}\,e^{-\sigma}\Big(
 2 T_{ae} \, G^{ec}\nabla_c\theta^{da}G_{db} 
 +e^{-\s}\theta^{ac} \nabla_c (e^\s T_{ab})   + \frac{e^{-\s}}{(n-1)} G_{bd} \theta^{cd} \nabla_c T  \Big) ,
 \label{eom-A-T}
\end{align}
which are valid for  arbitrary geometries. 
Note that \eq{eom-K-T} relates the mean extrinsic curvature 
$K^A = G^{ab} K_{ab}^A$ to the energy-momentum tensor.
As usual, the Maxwell-like equation \eq{eom-A-T} implies via the Bianci identity a wave equation
\begin{align}
 G^{ab}\nabla_a^{(g)}\nabla_b^{(g)} \theta^{-1}_{cd} &= \nabla_c J_d -
\nabla_c^{(g)} G^{ab}\nabla_b \theta^{-1}_{ad} - (c\leftrightarrow d) + \cO(R[g])
%+ R_{cade}[g] (g\theta^{-1}G)^{ea} + G^{ab}R_{cabe}[g]g^{ef}\theta^{-1}_{df} - (c\leftrightarrow d)
\label{maxwell-wave}
\end{align}
where $J_a$ is defined by the rhs of \eq{eom-A-T}.

\subsection{Perturbed flat branes and linearized gravity}
\label{sec:linearized}

Now recall the two possible interpretations of the matrix model:
A) as model for branes $\cM$ whose geometry is determined by $\phi^i$ resp. $F_{\mu\nu}$, as discussed above and
B) as a NC field theory on some given background, where $\phi^i$ resp. $F_{\mu\nu}$ are interpreted as 
$U(1)$-valued fields on a given background $\bar\cM$.
Let us now use B), which is more useful for the  perturbation theory, and
sufficient as long as the perturbations are small. As explained below, this can always be done  by
choosing a locally adapted background.

Thus consider some intrinsically flat
background solution $\bar x^A: \bar\cM\hookrightarrow \R^{10}$ of the bare matrix model without matter,
which satisfies $\bar\Box x^A = 0$.
We can furthermore assume  $\bar\nabla^{(g)} \bar\theta^{ab} = 0$ for the background solution
 without matter, since the intrinsic geometry is assumed to be flat;
this implies $\bar\nabla^{(g)} \bar G=0 = \del\bar\sigma$, 
hence $\bar\nabla^{(g)}=\bar\nabla^{(G)}\equiv \bar\nabla$.
We have in mind $\bar\cM = M^4 \times T^2$, as given explicitly in sector \ref{sec:tori}.
Now add  perturbations $\theta^{-1} = \bar\theta^{-1} + F, \ x^A = \bar x^A + \phi^A$.
To derive the equations of motion, we need the 
second order variation of the action \eq{action-semiclass} on this background $\bar\cM$ , given by
\begin{align}
 S^{(2)}[\phi,F]  &=  \frac{\L_0^4}{4(2\pi)^n}\!\! \int\!\!  d^{2n} x \sqrt{|G|} \Big(e^{\sigma} \bar G^{aa'} \bar G^{bb'} F_{ab}F_{a'b'}
 + 2 \bar  G^{ab} \del_a \phi_A \del_b \phi^A + 2 \phi^A m^2_{AB} \phi^B  \nn\\
 &  \qquad \qquad \quad + 4\bar \theta^{ac}F_{cd}\bar G^{db}  \phi_A K_{ab}^A \Big)   + S_{CS} 
\label{action-expanded-2}
\end{align}
where
\begin{align}
 m_{AB}^2 \ &= \ 2 K_{ab}^A \theta^{aa'}\theta^{bb'} K_{a'b'}^B  %\ \sim \ \frac{\theta^2}{r_{\cK}^2} ,
\label{scalar-mass}
\end{align}
using \eq{del-G-explicit}, and  $\l^a$ are Lagrangian multipliers 
which implement the constraint \eq{transversal-fluct}.
The ``would-be topological`` term\footnote{A simplified derivation of $S_{CS}$ could be given along the lines 
of section 4 in \cite{Blaschke:2010qj}.}  
\cite{Steinacker:2008ya} 
$S_{CS} \sim \int \rho\langle F\wedge F,\theta \wedge \hat\theta\rangle - \frac 12 (\hat\theta \to\eta\theta)$
is only relevant for the propagating gravitational modes considered in section \ref{sec:quadratic},
and can be dropped for $\bar\nabla\bar\theta^{ab} =0$.
The term
\begin{align}
S_{\rm mix} &= \frac{1}{(2\pi)^n} \int d^{2n}x\, \sqrt{G} \,\bar \theta^{ac}F_{cd}\bar G^{db}  \phi_A K_{ab}^A   \nn\\
 &= \frac{1}{(2\pi)^n} \int d^{2n}x\, \sqrt{G} \, T_{ab}[\phi] \,  \d_\cA  G^{ab} 
 = \frac{1}{(2\pi)^n} \int d^{2n}x\, \sqrt{G} \, T_{ab}[F] \,  \d_\phi  G^{ab}  
\label{S-mix}
\end{align}
couples the tangential and transversal perturbations. It can be written as 
a coupling to some effective induced energy-momentum tensors
\begin{align}
 T_{ab}[\phi] \, &= \ -\frac 12 \L_0^4\,\d_\phi  g_{ab}  =  \L_0^4\, \phi_A K_{ab}^A , \qquad 
  T_{ab}[\phi]\bar G^{ab} = 0   \nn\\
T_{ab}[F] \, &= \ \frac 14 \L_0^4\, F_{ad}\bar \theta^{db} g_{db} + (a \leftrightarrow b)
 + \frac 14 \L_0^4\, G_{ab} (F_{cd}\theta^{de}g_{ef}G^{fc})
\label{T-induced}
\end{align}
using the background equations of motion. We can then write  
 $S^{(2)}[\phi,F]$  as
\begin{align}
S^{(2)}[\phi,F] &= -\frac{\L_0^4}{(2\pi)^n} \int  d^{2n}x\,\sqrt{| G|} \,\Big(A_d\, \bar G^{da} \bar G^{bc}\bar \nabla_c(e^{\sigma} F_{ab})\, 
 + \phi^A\, (\eta_{AB}\bar\Box - m_{AB}^2) \phi^B  \Big) + S_{\rm mix}  \nn
%  &  +\frac{1}{(2\pi)^n} \int  d^{2n}x\,\sqrt{| G|} \, 
%  2 A_d \big (\bar  G^{db}\bar \theta^{ac} \bar \nabla_c T_{ab}[\phi]  + \bar \theta^{da}G^{bc}\, \bar\nabla_c T_{ab}[\phi] \big)
%\label{gravity-S-variation}
\end{align}
using \eq{dtildeG-form}.
Note that $\eta_{AB}$ will be positive  for the transversal fluctuations $\phi^A$, so there is no stability problem.
This gives the equations of motion 
\begin{align}
 \bar\Box\phi^A  + m_{CB}^2\eta^{AC} \phi^B \ &=  
\ - \L_0^{-4}\, e^{-\s}  K_{cd}^A \Pi^{cd}_{a'b'}  \bar\theta^{a'a} \bar\theta^{b'b} \ T_{ab}^{M+F}
 \label{eom-phi-T-2}\\
 \bar\nabla^b F_{ab}\ &=\  2 \L_0^{-4}\,e^{-\sigma}\Big(\theta^{bc} \bar\nabla_c  T_{ab}^{M+\phi}
 + G_{ad} \theta^{dc}\bar\nabla^b T_{cb}^{M+\phi}
 - \frac{1}{2n-2} G_{ad} \theta^{dc}\, \bar\nabla_c T^{M+\phi}  \Big)   \nn
% \label{eom-A-T-2}
\end{align}
where 
\begin{align}
 T_{ab}^{M+\phi} = T_{ab} + T_{ab}[\phi], \qquad \tilde T_{ab}^{M+F} = T_{ab} + T_{ab}[F] .
\end{align}
This implies again a wave-equation via \eq{maxwell-wave}
\begin{align}
 \bar\Box F = \nabla\nabla (T+T[\phi]) .
\end{align}
Given such a solution, we can  switch to the point of view B), and interpret these solutions 
in terms of a perturbed brane $\cM$ with deformed geometry. 
We can compute its curvature by 
inserting the solution into the expressions \eq{einstein-tensor-fluct} for the linearized Einstein tensor, 
noting the harmonic gauge\footnote{its validity in the presence of matter will be clarified later.} condition \eq{h-harmonic}.
This gives 
\begin{align}
\d\cG^{ab}  &= \d_\phi\cG^{ab}  \ + \ \d_F\cG^{ab}  \nn\\
&=  - e^{-\s}\bar\theta^{ac} \bar\theta^{bd} \bar K_{cd}^A \bar\Box \phi_A 
 - \frac 12 \bar\theta^{ac} \bar G^{bd} \bar\Box F_{cd}  - \frac 12\bar\theta^{bc} \bar G^{ad}\bar\Box F_{cd}  
 - \frac 14 \bar G^{ab} (\bar\theta^{cd}\bar\Box F_{cd})  \nn\\
 &= \cP^{ab;cd}\ T_{cd} \  +  \ \cO( \nabla\nabla (T+T[\phi])) ß + \cO((\phi,F)^2).
\label{einstein-equations-lin}
\end{align}
The quadratic terms $\cO((\phi,F)^2)$ are negligible compared with the linear ones, 
assuming that $K^A_{ab}$ is large for compactified branes.
As explained in the next section, one can always choose a locally adapted background with $F|_p=0$, hence 
the $T[F]$ was dropped for simplicity.
The tensor
\begin{align}
 \cP^{ab;cd} &= 
 \L_0^{-4}\, e^{-2\s}\bar\theta^{aa'} \bar\theta^{bb'} \bar K_{a'b';c''d''} \Pi^{c''d''}_{c'c'}\bar \theta^{c'c} \bar\theta^{d'd} 
\label{P-def}\\
  &=: G_N\; P^{ab}_{c'd'}\ \bar G^{c'c}\bar G^{d'd}  \    \nn\\[1ex]
G_N &=   \L_0^{-4}\, r_\cK^{-2} \ 
\label{G-newton-2n}
\end{align}
governs the coupling of $T_{ab}$ to the % ($\d_\phi\cG^{ab}$ contribution of the) 
 Einstein tensor. Here
\begin{align}
\bar K_{ab;cd} =  \bar\nabla_{a} \del_{b} \bar x_A\bar\nabla_{c} \del_{d} x^A \quad = \ \cO(r_\cK^{-2})
\end{align}
can be interpreted as extrinsic curvature of $\cM\subset \R^D$, as discussed
appendix A.
These equations amount to modified $2n$--dimensional linearized Einstein equations
\begin{align}
 \d\cG^{ab} &= G_N P^{ab}_{cd}\, T^{ab} \  + \  \cO(\nabla\nabla (T+T[\phi]))   \ .
\label{linearized-gravity}
\end{align}
$G_N$ plays the role of the effective $2n$-dimensional Newton constant, determined
by the extrinsic curvature scale $K_{ab;cd} \sim r_\cK^{-2}$ and the NC scale. It
will reduce to $G_N \sim \LNC^{-2}$ in 4 dimensions \eq{G-newton-4D}.

The main result is that the energy-momentum tensor $T^{ab}$ is coupled to the curvature through
the transversal perturbations $\phi^A$, leading to (Newtonian, at least) gravity.
The anisotropy of the coupling $\cP$ might be averaged out effectively in suitable compactifications\footnote{For example, 
all Lorentz-violating tensors such as $\langle \theta^{ab}\rangle = 0$ might be averaged out upon compactification,
reducing correlators such $\langle P^{ab}_{cd}\rangle \neq 0$ to their 4-D Lorentz-invariant averages.}.
On the other hand, $F$ determines perturbations of the orientation of the effective frame 
$\theta^{ab}\del_b$ \eq{frame}.
The derivative coupling of these curvature perturbations to $T^{ab}$ result from a 
coupling of $F$ to a dipole density $T_{ab}\theta^{bc}$  \cite{Steinacker:2010rh}, similar  as in electrodynamics. 
This can only lead to dipole and higher multipole  contributions to $F$, 
which is suppressed.

Moreover, the vacuum geometries are Ricci-flat provided the mixing contributions 
$T[\phi]$ and $T[F]$ vanish. 
We will indeed argue in section \ref{sec:quadratic} that for suitable backgrounds,
this mixing term  $S_{\rm mix}$ should lead to a splitting of the geometrical modes 
into massive ''optical`` modes which are irrelevant at
low energies, and massless Ricci-flat gravitational modes which do not mix.
In general however, the mixing may lead to a violation of  Ricci-flatness in vacuum.
This may be relevant for ''dark matter``, which at present is nothing but an
unexplained deviation from Ricci-flatness.
In any case we will basically ignore the mixing contributions in this paper,
leaving a detailed investigation for future work.

\subsection{Locally adapted backgrounds and gravity}
\label{sec:background}

We now study general brane  geometries at the non-linear level.
The  idea is to consider the space at any given point $p\in\cM$ as perturbation of some locally adapted, 
intrinsically flat (but not extrinsically flat) background $\bar\cM$ in terms of 
transversal and tangential perturbations
\begin{align}
x^A &= \bar x^A + \phi^A, \qquad
\theta^{-1}_{ab} = \bar\theta^{-1}_{ab} + F_{ab} , \nn\\ 
\phi^A|_p &= 0 = F|_p  ,
\label{perturb-normal} 
\end{align}
Since $\bar\cM$ is intrinsically flat,  $\theta^{ab}|_p$ can be extended on $\bar\cM$ such that
\begin{align}
 \bar\nabla^{(g)}\bar\theta^{-1}_{ab} &= \bar\nabla^{(g)} \bar G   = 0 = \del\bar\s
\end{align}
 implying that $\bar G$ is covariantly constant on $\bar\cM$,
so that $\bar\nabla^{(G)} =\bar\nabla^{(g)}\equiv \bar\nabla$.
It follows that the Laplacian on $\bar\cM$ is unique
\begin{align}
 \bar\Box \equiv \bar G^{ab}\bar\nabla_a\bar\nabla_b
\end{align}
$\bar\cM$ is the noncommutative analog of a  ''free-falling`` frame, exploiting the background independence of the 
matrix model. However, there are many possible backgrounds $\bar\cM\subset \R^{10}$ which are intrinsically flat but 
have different extrinsic geometry: one could simply choose the tangent plane $\bar\cM = T_p\cM$, or 
one can try to  match also the extrinsic curvature of $\cM$ with $\bar\cM$ by fitting
e.g. a cylinder or a cone. The latter is clearly more appropriate, because then a linear analysis of the 
corresponding perturbations $\phi_A$ suffices to compute the curvature perturbations, 
due to \eq{del-g-explicit}; this will become clear below.
Now the Gauss-Codazzi theorem $K_{ab;cd} - K_{bc;ad}  = R_{acbd}[g]$ tells us that we cannot expect to match 
the extrinsic curvature completely. However, in the case of compactified extra dimensions
such as  $\cM = \cM^4 \times T^2$, we can require that
\begin{align}
 \bar K_{ab}^A \approx K_{ab}^A  \qquad \mbox{and} \quad \bar K^A = \bar\Box\bar x^A = 0
\label{good-approx}
\end{align}
This implies that the intrinsic geometry of $\cM$ is nearly flat while the 
extrinsic curvature has large components, which is exactly satisfied for compactified branes.

We are now precisely in the situation of section \ref{sec:linearized} for linearized gravity. 
The perturbations $\phi^A, F$ viewed as  $U(1)$ perturbations on $\bar\cM$  must satisfy 
the equations of motion \eq{eom-phi-T-2}, since $\bar\cM$ is intrinsically flat. 
We can then switch to the point of view B), and interpret these solutions 
in terms of a deformed brane $\cM$ with perturbed geometry. 
Its linearized Einstein tensor is obtained as in \eq{einstein-equations-lin},
where we assume\footnote{cf. section \ref{sec:non-constant} for a discussion of the general case.} 
that $\bar\nabla \bar K_{ab}^A=0$ which is satisfied e.g. on cylinders.
This actually computes the full Einstein tensor at $p\in \cM$ since $\bar\cM$ is flat.
Since $p\in\cM$ was arbitrary,
we obtain the modified $2n$--dimensional Einstein equations 
\begin{align}
 \cG^{ab} &= G_N P^{ab}_{cd}\, T^{ab} \  + \  \cO(\nabla\nabla (T+T[\phi]))   \ + \cO((\phi,F)^2) \ .
\label{linearized-gravity}
\end{align}
where
\begin{align}
 \cP^{ab;cd} &= 
 \L_0^{-4}\, e^{-2\s}\theta^{aa'} \theta^{bb'} K_{a'b';c''d''} \Pi^{c''d''}_{c'c'} \theta^{c'c} \theta^{d'd} 
\label{P-def}\\
  &=: G_N\; P^{ab}_{c'd'}\  G^{c'c} G^{d'd}  \ ,  \nn\\[1ex]
G_N &=   \L_0^{-4}\, r_\cK^{-2} \ 
\label{G-newton}
\end{align}
Note that we tacitly replaced $\bar K_{ab}^A$ by $K_{ab}^A$, which requires \eq{good-approx} to hold.

To contrast the above considerations with general relativity, it is important to keep in mind that the 
fundamental degrees of freedom are very different.  
The fundamental degrees of freedom here are given by the transversal perturbations $\phi^A$ 
and the tangential perturbations $\cA^a$,
governed by the wave equation \eq{eom-phi-T-2}.
Those $\phi^A$  associated with extrinsic curvature $K_{ab}^A \neq 0$ couple linearly to $T$ and thus couple the geometry to matter, 
while the others couple only weakly or indirecty  to $T$ 
but nevertheless contribute to the geometry. The present mechanism seems unavoidable for 
compactified branes in matrix models, and must play a significant role for gravity on such branes.
If also the noncompact brane $\cM^4$ has extrinsic curvature, the non-linear terms $\cO((\phi,F)^2)$ may 
become significant and lead to 
embedding deformations called ``gravity bags'' \cite{Steinacker:2009mp}.
They may imply long-range modifications\footnote{The analysis in \cite{Steinacker:2009mp} is based on a 
somewhat different scenario using a complexified Poisson structure and needs to be adapted, but qualitative features are expected to 
carry over to the present framework.} of gravity, possibly relevant to galactic or cosmological scales.

The bottom line is that the energy-momentum tensor indeed couples  to the Einstein tensor
 in the ``semi-classical'' matrix model,  without invoking any quantum effects or induced
Einstein-Hilbert terms. 
The transversal brane fluctuations provide the coupling of gravity to matter, and
complement the tangential $F$ modes first observed in \cite{Rivelles:2002ez}.
Vacuum geometries are Ricci-flat if the mixing contributions $T[\phi], T[F]$ drop out
and the non-linear terms $\cO((\phi,F)^2)$ are negligible.
This link between curvature and the e-m tensor was missing in the earlier related works
 \cite{Rivelles:2002ez,Steinacker:2009mp,Yang:2010kj}, finally providing   
a possibly realistic mechanism of emergent gravity scenario in matrix models.

Up to now, we have been studying  the $2n$-dimensional geometry of $\cM$.
In order to understand the effective 4-dimensional gravity, we consider in the next section
compactified backgrounds  in more detail. 
Then we will indeed obtain 4-dimensional gravity (at least for toroidal compactifications), and identify 
the effective 4-dimensional Newton constant.

\section{Compactified branes and 4-dimensional gravity}
\label{sec:compact-split}

We have seen that the coupling of gravity to matter requires the presence of extrinsic curvature.
Let us therefore discuss in more detail branes with compactified extra dimensions
\be
\cM^{2n}  = \cM^4 \times \cK \quad \subset \ \R^D
\label{compactification}
\ee
where the extrinsic curvature is predominantly due to $\cK \subset \R^{D}$, while the embedding
of $\cM^4$ is approximately flat.
Such solutions for\footnote{The intrinsic geometry of the solutions denoted $S^2 \times T^2$ and $T^4$
 in  \cite{Steinacker:2011wb} is in fact $S^3\times S^1$.} $\cK = T^2, \ \cK = S^2 \times S^2$ and $\cK = S^3 \times S^1$ 
were given recently \cite{Steinacker:2011wb}. 
While the induced metric $g_{ab}$ on $\cK$ is space-like, the effective metric 
$G_{ab}$ on $\cK$ is degenerate or has Minkowski signature, corresponding to light-like compactification. 
This is possible because of  "split noncommutativity``, where the
Poisson bi-vector  relates the compact space $\cM^4$ with the non-compact space $\cK$,
\be
\theta^{ab}\del_a \wedge \del_b =  \theta^{\mu i}(x,y) \frac{\del}{\del x^\mu} \wedge \frac{\del}{\del y^i}\quad + ... \ 
\label{split-NC}
\ee 
where $x^\mu$ are coordinates on $\cM^4$ and $y^i$ are coordinates on $\cK$.
Such a structure is realized e.g. by the canonical symplectic form on the cotangent bundle $T^* \cK$.
If $\cK$ has dimension 4, then $\cM^4$ might even be isotropic, $\{x^\mu,x^\nu\} =0$. 
Now recall that metric variation due to embedding fluctuations is given by
 $\d g_{ab} = -2\phi_A K_{ab}^A$  \eq{del-g-explicit}.
For the present type of background, this implies that only the
perturbations $\phi_A$ of the compactification $\cK$ couple to matter,
while the perturbations of flat $M^4 \subset \R^4$ decouple.
Remarkably, such perturbations of $\cK$
lead to perturbations of the effective 4-dimensional (!) metric on $M^4$  due to split noncommutativity,
\begin{align}
\d_\cK\big(\g^{ab} \del_a\otimes \del_b\big)  \ &\approx \ \theta^{a i} \theta^{b j} \d g_{i j}^\cK\, \del_a\otimes \del_b
 \ \equiv \  \d \g^{ab}_\cK \,  \del_a\otimes \del_b \  
\approx \d  \g^{\mu\nu}_\cK \, \del_\mu\otimes \del_\nu \,
\label{dG-K}
\end{align}
 assuming $\theta^{ij} \approx 0$, in self-explanatory notation. This will be elaborated in  detail below.

\paragraph{Constant curvature compactifications and moduli.}

To make this more explicit, we assume that 
\be
\cK = \times_i\, \cK_{(i)}\ \subset \R^D
\label{product-compact}
\ee
is a product manifold with
constant exterior curvature, 
in the sense that
\begin{align}
 \nabla_a\nabla_b \bar x^A &= K_{ab}^A\  =\ - \sum_i \frac{1}{r_i^2}\,  g_{ab}^{(i)}\, \bar x^A_{(i)} .
\label{nice-geom-cond-1}
\end{align}
Here $g_{ab}^{(i)}$ is the induced metric on $\cK_i$ with radius $r_i$.
This holds e.g. for  $\cK = T^n = \times_i S^1_{(i)}$ or $\cK = S^n$,
which suffices to understand the mechanism. 
Following the discussion in section \ref{sec:background}, we choose at any given point $p\in\cM^4$
a locally adapted intrinsically flat background cylinder (or cone) $T_p\cM^4 \times \cK$ with constant radii $\bar r_i$
and $\bar\nabla\bar\theta=0$.
Perturbing the radii and $F$ such that $\d r_i|_p = 0 = F|_p$ leads to
\begin{align}
\d g_{ab} &= 2 \sum_i \frac{1}{r_i}\,\d r_i \,g_{ab}^{(i)} ,
\label{dg-compact}
\end{align}
where
\begin{align}
\d r_i = \frac 1{r_i}\,\phi_A^{(i)} \bar x^A_{(i)}  % = \d r_i(x)
\end{align}
denotes the radial moduli of the  $\cK_i$, which play the central role in the following.
Then 
\begin{align}
\d\gamma^{ab} &= 2 \sum_i \frac{1}{r_i}\,\d r_i \,\bar \g^{ab}_{(i)} \ + \cO(F) \nn\\
\d\s &=  \frac 1{n-1} \sum_i \frac{1}{r_i}\,\d r_i \,g_{ab}^{(i)} g^{ab}_{(i)} \ + \cO(F) \
= \ \frac 1{n-1} \sum_i\dim \cK_i\, \frac{1}{r_i}\,\d r_i \, \ + \cO(F)
\end{align}
where
\begin{align}
 \g^{ab}_{(i)} \ \equiv \   \g^{ab}_{\cK_i} = \theta^{a k} \theta^{b l}  g_{kl}^{(i)}.
\label{gamma-const-curv}
\end{align}
Then 
\begin{align}
\bar \Box \d\gamma^{ab} &= 2 \sum_i \frac{1}{r_i}\,\bar \Box\d r_i \,\g^{ab}_{(i)} \ + \cO(F) . \label{box-gamma-const}
\end{align}
using $\bar \nabla \bar \g^{ab}_{(i)} = 0$ by the above assumptions.
The equations of motion for these radial moduli $\d r_i$ are obtained from \eq{eom-phi-T-2} 
\begin{align}
 \bar\Box \d r_i \ &=  \frac{\L_0^{-4}}{r_i}\, e^{-\s} T_{ab}\, \bar\theta^{aa'} \bar\theta^{bb'} \Pi^{cd}_{a'b'} g_{cd}^{(i)}  
 =  \frac{\L_0^{-4}}{r_i}\, T_{ab} \, e^{-\s}\Big(\g^{ab}_{(i)} - \frac{\dim \cK_i}{2(n-1)}\,\g^{ab} \, \Big)   .
\label{eom-dr}
\end{align}
while the story for $F$ is as before and will not be repeated.
Here $g_{cd}\,g^{cd}_{(i)} = g_{cd}^{(i)}\,g^{cd}_{(i)} = \dim \cK_i$,
since the embedding of $\cK_i$ is assumed to be orthogonal 
to $\cM^4$ (and possible other $\cK_j)$. 
Switching  to the geometrical picture, these perturbations correspond to metric perturbations with
 $2n$--dimensional Einstein tensor  \eq{einstein-tensor-fluct}
\begin{align}
\cG^{ab} &=  e^{-\s} \sum_i \frac 1{r_i} \bar\g^{ab}_{(i)} \bar\Box_G \d r_i   
 - \frac 12  \bar G^{bd} \bar\theta^{ac} \bar\Box F_{cd}  - \frac 12\bar G^{ad} \bar\theta^{bc} \bar\Box F_{cd}
 + \frac 14 \bar G^{ab} (\bar\theta^{cd} \bar\Box F_{cd})  +  \cO(\d^2) \nn\\
 &= \cP^{ab;cd}\ T_{cd}\ +   \ \cO(\L_0^{-4} \nabla\nabla (T+T[\phi])) + \cO(\d^2) .
\label{einstein-equations-moduli}
% \label{cG-eff-fluct-K}
\end{align}
This holds for toroidal compactifications $M^4 \times T^{2m}$, 
and $\cO(\d^2)$ stands for quadratic contributions in the perturbations.
Here $\cP$ is given  by
\begin{align}
 \cP^{ab;cd}&= 
 \L_0^{-4}\, e^{-2\s} \sum_i \frac 1{r_i^2}\,\g^{ab}_{(i)}\Big(\g^{cd}_{(i)} - \frac{\dim \cK_i}{2(n-1)}\,\g^{cd} \,\Big) .
\end{align}

\subsection{Effective 4-dimensional gravity for toroidal compactifications}

To obtain the 4-dimensional Einstein tensor, we simply perform a dimensional reduction along $\cK$.
This leads to the effective 4-dimensional metric as derived more generally in section \ref{sec:generalizations},
\begin{align}
 G_{(4)}^{\mu\nu} =  e^{-\s_4}  \gamma^{\mu\nu}
\end{align}
where the normalization $e^{-\s_4}$ plays the same role 
in 4 dimensions as $e^{-\s}$ does on $\cM^{2n}$.
Then the  $2n$-dimensional Laplacian
$\Box$ can  be related to the 4-dimensional Laplacian as follows (cf. \eq{harmon-reduced})
\begin{align}
 \Box_{(4)} = G_{(4)}^{\mu\nu} \nabla_\mu\nabla_\mu  \ =  \  e^{-\s_4 + \s}\Box
\end{align}
if acting on tensor fields that are covariantly constant on $\cK$.
We can then repeat the above
computation leading to \eq{einstein-equations-moduli} for $G_{(4)}^{\mu\nu}$ on $M^4$,
replacing $\Box_{(4)} \d r_i $ by $\Box \d r_i$ and subsequently using  \eq{eom-dr}.
The harmonic gauge condition still applies in 4 dimensions \eq{h-harmonic-4d}, and 
we obtain the following equation for the effective 4-dimensional Einstein tensor 
\begin{align}
\cG_{(4)}^{\mu\nu} 
= \ \cP^{\mu\nu;cd}_{(4)}\ T_{cd}\ +   \ \cO(\L_0^{-4} \nabla \nabla (T+T[\phi])) + \cO(\d^2) 
 \label{einstein-equations-4D}
\end{align}
where\footnote{Notice that  \eq{einstein-equations-moduli} makes sense as tensorial equation both on $\cM^{(2n)}$ and on $M^4$.  
However, the $2n$- dimensional metric couples $M^4$ with $\cK$, and any other form of \eq{einstein-equations-moduli}
e.g. in terms of $T^{ab}$ would not restrict to $M^4$.}
\begin{align}
\cP^{\mu\nu;cd}_{(4)} &= 
 \L_0^{-4}\, e^{-2\s_4} \sum_i \frac 1{r_i^2}\,\g^{\mu\nu}_{(i)}\Big(\g^{cd}_{(i)} - \frac{\dim \cK_i}{2(n-1)}\, \g^{cd} \,\Big) .
\label{P-4D}
\end{align}
This is similar to the 4-dimensional Einstein equations.
In particular, these metrics are  Ricci-flat in vacuum, apart from possible mixing 
contributions $T[\phi]$ and nonlinear effects $\cO(\d^2)$.
We keep the $2n$-dimensional $T_{ab}$ for the sake of generality,
but for most applications the e-m tensor will be 4-dimensional.
The effective 4-dimensional Newton constant or the Planck length 
is determined by the compactification scale $r_\cK^{-2}$ as well as $\LNC$ analogous as in \eq{G-newton},
\begin{align}
G_N =  \L_0^{-4} r_\cK^{-2} \  = \ g_{\YM}^2\, e^{\s_4} \,r_\cK^{-2} \ .
\label{Newton-const-4d}
\end{align}
Here 
\begin{align}
\frac 1{g_{\YM,4}^2} = \L_0^{4} e^{\s_4} 
\label{gauge-coupling-4d}
\end{align}
is the effective four-dimensional gauge coupling\footnote{We are considering the action at the classical level here.
In a more complete treatment, these constants will receive quantum 
corrections, and the action should be replaced by the low-energy quantum effective action
since gravity is a low-energy phenomenon. Some protection from 
quantum corrections is expected due to supersymmetry of the matrix model.}
in analogy to \eq{gauge-coupling}, and
$e^{-\s_4}$ is related to the noncommutativity scale.
This will be elaborated in more detail for $M^4 \times T^2$ below.

We conclude that the compactification moduli of $\cK$ describe perturbations of the 
 effective 4-dimensional metric $G^{\mu\nu}_{(4)}$, and encode gravitational degrees of freedom. 
Since they couple linearly to the energy-momentum tensor, Newtonian gravity
is recovered, as elaborated below.
This is a remarkable mechanism, which offers also an unexpected solution of the moduli 
stabilization problem in matrix models. 
%However, some of these moduli may become massive due to  fluxes:

\paragraph{Flux compactifications and moduli stabilization.}

In the presence of fluxes on $\cK$, some of the embedding moduli of $\cK\subset \R^{10}$ are 
stabilized by the effective mass $m_{AB}^2$ \eq{scalar-mass}.
To see this, recall the quadratic action \eq{sec:quadratic} in the case of the product compactification as above.
Then
\begin{align}
m_{AB}^2 \phi^A \phi^B\ &= \ 2\sum_{i,j}\,g_{ab}^{(i)} \g_{(j)}^{ab} \
 \frac{\d r_i}{r_i}\, \frac{\d r_j}{r_j} .
%m^2_{AB}\phi^A \phi^B = \d_{AB}\phi^A \phi^B\frac 1{R^2}\, \d_{ij} \d_{i'j'} \theta^{ii'}\theta^{jj'}\, 
\end{align}
This means that the radial moduli $\d r_i$  are long-range propagating modes (gravitons) in the absence of flux,
but may acquire a mass in the presence of a flux on $\cK_i$.
Note that the flux $\theta^{ab}$ may connect different $\cK_i$, 
which happens e.g. for fuzzy tori $T^2_N \sim S^1 \times S^1 \subset \R^4$.
In that case the mass term 
\begin{align}
m^2_{AB}\phi^A \phi^B = 2 \theta^2_{(12)}\, \frac{\d r_1}{r_1}\, \frac{\d r_2}{r_2}
\end{align}
has indefinite signature, so that
compactifications on tori with fluxes are unstable. Therefore fuzzy cylinders 
leading to tori without fluxes \cite{Steinacker:2011wb} are better. 
Then the massless modes $\d r_i$ are the radial modes of the cycles, which may vary along $\cM^4$. 
More generally, we can 
 diagonalize $m^2_{AB}$ by some $x$-dependent $SO(D-4)$ transformation, 
\begin{align}
 m^2_{AB} = \oplus_i\ m^2_{(i)}\,\d_{AB}^{(i)} .
\label{moduli}
\end{align}
The condition $m^2_{(i)}=0$ determines massless moduli fields $\d r_i$ among the transversal perturbations $\phi^A$. 
Since they do not couple to a flux on $\cK$,
the fluctuations of the  metric due to these massless moduli are  
along the non-compact  $M^4$ as in \eq{dG-K}, due to split noncommutativity
\be
\d\g^{ab}_{(i)} \del_a\otimes \del_b\ = \ \d\g^{\mu\nu}_{(i)} \del_\mu\otimes \del_\nu\ .
\label{dG-K-tang}
\ee
Note that the above mass term also applies to the nonabelian scalars. Therefore a flux on $\cK$
 typically leads to SUSY breaking, which is  well-known in string theory 
(see e.g. \cite{Kachru:2002he}).
If $\cK$ is 4-dimensional, then there should be no flux on $\cK$ if the two
transversal degrees of freedom are to remain massless. 
These issues are discussed further in section \ref{sec:generalizations}.

\subsection{Gravitational excitations.}
\label{sec:quadratic}

Let us briefly  discuss the  geometrical modes 
(i.e. analogs of gravitons) in vacuum,
assuming $\cM^{2n} = \cM^4 \times \cK \subset \R^{9,1}$. 
 There are $10-2=8$ physical degrees of freedom in the $U(1)$ sector of the IKKT model,
 taking into account gauge invariance.
We can distinguish three different types of modes: 
\begin{enumerate}
 \item  transversal (i.e. radial) modes $\phi^A$ on $\cK$ corresponding to extrinsic curvature.
They are clearly relevant to gravity since they couple to matter as discussed above. 
Only the massless moduli $\phi^A_{(\a)}$ \eq{moduli}  are relevant at low energy.
\item
the remaining transversal modes $\phi^A$ on $\cM^4$  not corresponding to extrinsic curvature directions. 
These are completely sterile at the linearized level.
They may play a role in long-range modifications of modify gravity if  $\cM^4$ has weak extrinsic curvature, 
 describing ``gravity bags'' \cite{Steinacker:2009mp}.
 \item
 the $2n-2$ tangential modes $F_{ab}$ corresponding to would-be $U(1)$ gauge fields.
\end{enumerate}

However, some of these would-be massless modes are coupled via $S_{\rm mix}$ \eq{S-mix},
which involves the compactification scale. 
Such a mixing typically modifies the dispersion relation, and we expect that those
modes which participate in the mixing will become massive or ``optical'', thus dropping
out from low-energy physics. Since the mixing is mediated by the radial moduli $\d r_i$, 
$2l$ of these modes should become massive in this manner, where $l$ is the number of radial moduli.
The remaining  modes which do not participate in the mixing remain massless, and therefore 
lead to Ricci-flat (!) metric perturbations \eq{einstein-equations-moduli}.
Hence for $l=2$ (e.g. for $\cK=T^2$ or $\cK = S^1 \times S^3$), 
we should be left with $8-4=4$ massless modes. 
Two of these could play the role of physical gravitons
(recall that the gravitons arise automatically in harmonic gauge), and the remaining two will be geometrically trivial 
and might be absorbed in the scalar fields 
$e^{-\s}$  and $\eta \sim G^{ab}g_{ab}$, which are somewhat reminiscent of dilaton and axion.
Of course this rather appealing picture is based on  assumptions which are not yet justified, and 
clearly depends on the compactification.
A more detailed study of these modes will be given elsewhere.

\subsection{Explicit example: $M^4 \times T^2$}
\label{sec:tori}

Let us work out the above results explicitly for the case of compactifications on $T^2$,
\be
\cM = \R^4 \times T^2 \quad \subset \R^D .
\ee 
We can  assume that the non-compact  $\R^4$ is embedded along the $0,1,2,3$ directions.
To obtain $\R^4 \times T^2  \ \subset \R^8$, we start with
two fuzzy cylinders $(U_4,X^2)$ and $(U_5,X^3)$ with NC parameter $\kappa_{(4,5)}$ and radii $r_{4,5}$ 
defined via
\begin{align}
 [U_4,X^2] &= \kappa_4 U_4,  \qquad [U_5,X^3] = \kappa_5 U_5,  \nn\\ 
U_i U_i^\dagger &= r_i^2, \qquad i = 4,5 \ .
\end{align}
We can make them rotate along a two-dimensional {\nc} plane $[X^\mu,X^\nu] = i \theta^{\mu\nu}, \,\mu = 0,1$
(which commutes with the cylinders) as follows \cite{Steinacker:2011wb}
\begin{align}
X^A = \begin{pmatrix}
 X^{0,1} \\
 X^2 \\
 X^3 \\
 {X^4} + i {X^5}\\
{X^6} + i {X^7}
\end{pmatrix}
=
\begin{pmatrix}
 X^\mu \\
U_4\, e^{i k^{(4)}_\mu X^\mu} \\
U_5\, e^{i k^{(5)}_\mu X^\mu}
\end{pmatrix} .
\label{R4T2-solution}
\end{align}
These are solution of the matrix equations of motion 
\be
\Box X^A = 0\qquad \mbox{if} \qquad  
k_\mu^{(i)}  k_\nu^{(i)} \theta^{\mu \mu'} \theta^{\nu \nu'} \eta_{\mu'\nu'}   = -  \kappa_i^2  \quad (\mbox{no sum over}\ i ) ,
\ee
provided  $[k^{(4)}_\mu X^\mu,k^{(5)}_\nu X^\nu] = 0$.
These solutions describe $\R^4\times T^2$ where the torus rotates along the non-compact space, and is stabilized by 
angular momentum.
Note that the only non-vanishing  4-dimensional component of $\theta^{\mu\nu}$ is $\theta^{01} \neq 0$, where $x^0$
is time-like w.r.t. $g_{ab}$. This is essential to obtain an effective 4-dimensional
metric with Minkowski signature, as we will see. Moreover, this 6-dimensional solution  
behaves as a 4-dimensional (!) space 
$\R^2 \times T^2$ in the UV  \cite{Steinacker:2011wb}, such that the IKKT model is (expected to be)
perturbatively finite on this background.

\paragraph{Semi-classical analysis.}

To gain more insights into this solution and its effective metric, we consider the
semi-classical limit. Then the above solution can be described  in terms of 
a 6-dimensional plane compactified on a 2-torus. Consider 6-dimensional coordinates 
 $\xi^a = (x^\mu,\xi^4,\xi^5)$, and $U_4 \sim r_4 e^{i \zeta^4}$ 
and $U_5 \sim r_5 e^{i \zeta^5}$.
Then  \eq{R4T2-solution} can be written in a compact way as
\begin{align}
x^A \ \ = \ \  \begin{pmatrix}
       x^\mu \\ r_4\, \exp(i (k^{(4)}_\mu x^\mu + \zeta^4)) \\
              r_5\, \exp(i (k^{(5)}_\mu x^\mu + \zeta^5)) 
      \end{pmatrix}
\ \ \equiv \ \  \begin{pmatrix}
            x^\mu \\ r_4\, \exp(i \xi^4) \\
              r_5\, \exp(i \xi^5) 
      \end{pmatrix}
\label{R4T2-solution-short}
\end{align}
(dropping constant phase shifts). The tori are compactified along the 6-dimensional momenta
\be
k^{(4)} = (k_{(4)}^\mu,1,0), \quad k^{(5)} = (k_{(5)}^\mu,0,1) 
\ee
in the $(x^\mu,\zeta^i)$ coordinates, or along $\xi^4, \xi^5$ in the $\xi^a$ coordinates; the 
latter are more useful here.
The Poisson tensor can be written as 
\begin{align}
\theta^{ab} =  \{\xi^a,\xi^b\} = \theta \begin{pmatrix}
0 & c &  0 & 0  & \vartheta^{0}_4 & \vartheta^{0}_5\\
-c & 0 & 0 & 0  & \vartheta^{1}_4& \vartheta^{1}_5\\
0 & 0 & 0 & 0 & 1& 0\\
0 & 0 & 0 & 0& 0 & 1 \\
-\vartheta^{0}_4 & -\vartheta^{1}_4 &-1 &  0 & 0 & 0 \\
-\vartheta^{0}_5 & -\vartheta^{1}_5 & 0 &-1 & 0 &0
 \end{pmatrix} .
\label{theta-6D}
\end{align}
Of course we could admit  more general $\vartheta^{\mu}_{4,5}$.
Here the no-flux condition $\theta^{45} = 0$ is already imposed,
which amounts to\footnote{this leads to a constraint on $k^{(4)}_3$ and $k^{(5)}_2$ in \eq{R4T2-solution-short}.}
\be
[z^4,z^5] = 0,\qquad z^4 = x^4+ix^5, \quad z^5 = x^6+ix^7.
\ee
The embedding metric  is obviously flat, given by
\begin{align}
g_{ab} &= (\eta_{\mu\nu},4\pi^2 r_i^2\, \d_{ij}) = \diag(-1,1,1,1,4\pi^2 r_4^2,4\pi^2 r_5^2)
\end{align}
which is diagonal in the $\xi^a$ coordinates.
Therefore $e^{(a)} = \theta^{ab}\del_b$ defines a frame \eq{frame} of orthogonal (but not orthonormal) tangent vectors
on $\R^4 \times T^2$, which satisfy
\begin{align}
 (e^{(a)},e^{(b)})_G = \theta^{aa'}\theta^{bb'}G_{a'b'} = e^\s g^{ab}= e^\s(\eta^{\mu\nu},\frac 1{4\pi^2 r_i^2} \, \d_{ij}) .
\end{align}
Note that the effectively time-like vector $e^{(0)} = c\theta\del_1 + ...$ 
is pointing along $x^1$ (rather than $x^0$!) and is wrapping the torus.
The  6-dimensional conformal metric is given in $\xi^a$ coordinates by 
\begin{align}
\g^{ab} &= \theta^{aa'}\theta^{bb'} g_{a'b'} = \theta^2\left(\scriptsize
\begin{array}{c|c}
 \theta^{-2}\g_{(4-d)}^{\mu\nu}   & \begin{matrix}
                         -c \vartheta^1_4 & -c \vartheta^1_5 \\
                         -c \vartheta^0_4 & -c \vartheta^0_5 \\
                              0 & 0 \\
                             0 & 0   \end{matrix}  \\
 \hline  \\
\begin{matrix}
  -c \vartheta^1_4 & -c \vartheta^0_4 & 0 & 0 \\
  -c \vartheta^1_5 & -c \vartheta^0_5 & 0 & 0 
\end{matrix}  & 
\begin{matrix}
 & 1-(\vartheta^0_4)^2+(\vartheta^1_4)^2 & -\vartheta^0_4 \vartheta^0_5+\vartheta^1_4 \vartheta^1_5 \\
 & -\vartheta^{0}_4 \vartheta^0_5+\vartheta^1_4 \vartheta^1_5 & 1-(\vartheta^0_5)^2+(\vartheta^1_5)^2
\end{matrix}
\end{array}
\right) . \nn
\end{align}
In particular the  metric $G^{ab}$ of $\R^4\times T^2$ is also flat, but 
it is not a product metric: $T^2$ is not perpendicular to $\R^4$.
This must be so, because the compactification must be time-like {\em and} the 4-dimensional
space must have Minkowski signature.
However, all we need for 4-dimensional physics is the 
 4-dimensional metric, which  is given by 
\begin{align}
G_{(4)}^{\mu\nu} \ &= \ e^{\s_4}\,\g^{\mu\nu}_{(4-d)} \equiv e^{\s_4}\,\g^{\mu\nu},  \nn\\
\g^{\mu\nu}\  &= \ \theta^{\mu a}\theta^{\nu b} g_{ab} 
=\theta^2\, \diag(c^2,-c^2,0,0) + e^{(4)\mu}e^{(4)\nu}\, 4\pi^2 r_4^2 \, + e^{(5)\mu}e^{(5)\nu}\, 4\pi^2 r_5^2 \, \nn\\
\ &= \  - e^{(0)\mu}e^{(0)\nu}\, c^2 + e^{(1)\mu}e^{(1)\nu}\, c^2
  + e^{(4)\mu}e^{(4)\nu}\, 4\pi^2 r_4^2 \, + e^{(5)\mu}e^{(5)\nu}\, 4\pi^2 r_5^2 \,  
\label{G-frame-4D}
\end{align}
which is clearly non-degenerate with Minkowski signature. 
The $e^{(i)}$ for $i=0,1,4,5$ form an orthogonal (but not orthonormal) frame for the 
effective 4-dimensional metric, where
$e^{(0)}$ is time-like and the others are space-like.  
Explicitly the 4-dimensional frame is
\begin{align}
 e^{(4)\mu} &= \theta(\vartheta^{\mu}_4 + \d^{\mu}_2), \qquad  e^{(1)\mu} = \theta\d^{\mu}_0  \nn\\
 e^{(5)\mu} &= \theta(\vartheta^{\mu}_5 + \d^{\mu}_3)  , \qquad  e^{(0)\mu} = \theta\d^{\mu}_1
\end{align}
where $e^{(4,5)a} \neq 0$ only for 4-dimensional indices $a \to \mu =0,1,2,3$. 
The equations of motion  $\Box z^j=0$ reduce to
\begin{align}
0 &= G^{44} \ \propto\  1-(\vartheta^0_4)^2 + (\vartheta^{1}_4)^2 ,  \nn\\
0 &= G^{55} \ \propto \ 1-(\vartheta^{0}_5)^2+(\vartheta^{1}_5)^2 ,
\label{torus-onshell}
\end{align}
expressing the fact that the compactification is light-like.
A possible solution is 
\begin{align}
 \vartheta^0_4 &= 1 = \vartheta^0_5, \qquad  \vartheta^{1}_4 = 0=  \vartheta^{1}_5  \nn\\
e^{(4)\mu} &= \theta(1,0,1,0), \qquad e^{(5)\mu} = \theta(1,0,0,1) .
\end{align}
The  conformal factors are determined by \eq{g-4-normalization}
\begin{align}
 e^{-\s}\, \sqrt{|G_{ab}|} &=\sqrt{{|\th^{-1}_{ab}|}}  = \frac{e^{-\s_4}}{V_0} \,\sqrt{|G^{(4)}_{\mu\nu}|}\, , \nn\\ 
 e^{\s_4}  % &= V_0 \frac{\sqrt{\th^{-1}_{ab}}}{\sqrt{\det{\g^{(4)}_{\mu\nu}}}} 
 &=   V_0 \sqrt{{|\th^{-1}_{ab}|}|{\g^{(4)}_{\mu\nu}|}^{-1}} 
\label{sigma4-explicit-torus}
\end{align}
where $V_0 = \int_{T^2} d\xi^4 d\xi^5 = (2\pi)^2$ in the $\xi^a$ coordinates.
%and $\det \theta^{ab} = \theta^6 c^2$.

\paragraph{Metric fluctuations.}

Now we can illustrate the mechanism for gravity. Consider  transversal fluctuations $\phi^A = \d x^A$
around the above toroidal background $z^i = r_i \exp{i \xi_i},\ i=4,5$  as in section \ref{sec:perturbations}.
We will use a linearized approach here, and  omit the tangential perturbations $F$ for simplicity. 
Then the transversality  constraint \eq{transversal-fluct} is identically satisfied 
by the ansatz 
\be
\d z^i = \phi^i = \, \frac{z^i}{r_i}\,\d r_i, \qquad i=4,5 
\ee 
in complex notation.
These radial fluctuations $\d r_i$ lead to  metric fluctuations 
\begin{align}
\d g_{ii} &= -2\phi_A K_{ii}^A = 8\pi^2 r_i\d r_i\, , \qquad i=4,5 \quad \mbox{(no sum)}, \nn\\
 \d \g^{ab} &\to  \d \g^{\mu\nu} = \sum_{i=4,5} \d \g^{\mu\nu}_{(i)} 
  = 8\pi^2\Big( e^\mu_{(4)}e^\nu_{(4)}\, r_4\d r_4 \, + e^\mu_{(5)}e^\nu_{(5)}\, r_5\d r_5 \,\Big) \nn\\
 \d  G^{ab} &= e^{-\s} \d \g^{ab}  -  G^{ab} \d\s  
\end{align}
where
\begin{align}
  \d \s &= \frac 14\, g^{ab} \d g_{ab}  = \frac 12\, \big( r_4^{-1}\d r_4 + r_5^{-1}\d r_5 \big) 
\end{align}
Note again that $\d \g^{ab}\del_a\otimes\del_b   = \d \g^{\mu\nu}\del_\mu\otimes\del_\nu$ affects only the 4-dimensional metric on $\R^4$,
in accordance with \eq{dG-K}. Hence the $\d r_i$
become two space-like metric degrees of freedom \eq{G-frame-4D},
 governed by the effective action \eq{action-expanded-2}, \eq{dS-matter}
\begin{align}
S & \sim \int_{\cM^6} \sqrt{|G|}\, \big(\L_0^4\, G^{ab}\del_a\phi^A \del_b \phi_A + \frac 12 T_{ab}\d  G^{ab} \big) \nn\\
 &= \int_{M^4} \sqrt{| G_{(4)}|}\, \Big(\L_0^4\,\sum_{i=4,5}\, G_{(4)}^{\mu \nu} \del_\mu\d r_i \del_\nu\d r_i
 + T_{\mu\nu} e^\mu_{(i)}e^\nu_{(i)}\,4\pi^2 e^{-\s_4} r_i\d r_i \ 
 -\frac 12  T_{\mu\nu}\g^{\mu\nu} e^{-\s_4} r_i^{-1}\d r_i  \Big) \nn\\
 &= \int_{M^4} \sqrt{| G_{(4)}|}\, \big( \L_0^4\, \sum_{i=4,5}\,G_{(4)}^{\mu \nu} \del_\mu\d r_i \del_\nu\d r_i 
 + \frac 12 T_{\mu\nu}\d  G_{(4)}^{\mu\nu}  \big) 
\label{S-torux-explicit}
\end{align}
dropping the mixing contributions $S_{\rm mix}$ for simplicity.
Here we note that
\begin{align}
G^{ab} \del_a \d r\,(z^i, \del_b z^i)_g\ &= 0 
\end{align}
using the on-shell condition $G^{44} = G^{55} = 0$, and furthermore
$m^2_{AB} = 0$ since there is no flux on $T^2$.
We restrict ourselves to the lowest KK modes $\d r_i = \d r_i(x^\mu)$, and assume
$T^{\mu\nu}$ is 4-dimensional.
The perturbation of the effective 4-dimensional metric is\footnote{The attentive reader may notice an apparent
mismatch between the trace contributions from the $2n$- and the 4-dimensional point of view in \eq{S-torux-explicit},
which will be resolved in the next section.}
\begin{align}
 \d  G^{\mu\nu}_{(4)} \ &= \  e^{-\s_4} \d \g^{\mu\nu}  -  G^{\mu\nu}_{(4)} \d\s_4 \nn\\
\d\s_4 \ &= \  \frac 12 \g_{\mu\nu}\d \g^{\mu\nu} =  r_4^{-1}\d r_4 + r_5^{-1}\d r_5 
\label{dGs-torus}
\end{align}
using \eq{sigma4-explicit-torus}.
 Therefore we obtain the equations of motion for $\d r_i$ 
\begin{align}
\L_0^{4} \Box_{(4)} \d r_i &=T_{\mu\nu} e^\mu_{(i)}e^\nu_{(i)}\, 4\pi^2 \ e^{-\s_4} r_i \ 
- T_{\mu\nu}\g^{\mu\nu} e^{-\s_4} \frac 1{2r_i}
 = T_{\mu\nu} e^{-\s_4}\frac 1{r_i}\(\g^{\mu\nu}_{(i)} - \frac 12\,\g^{\rho\eta} \,\) .
\label{dr-eqn-t2}
\end{align}
Note that the $\d r_i(x)$ are indeed massless moduli, reflecting the fact that
the on-shell condition \eq{torus-onshell}
is independent of the radii $r_i$.
This leads to the following equation for the  4-dimensional linearized Einstein tensor 
\begin{align}
\d_\phi\cG^{\mu\nu}_{(4)}  &=  e^{-\s_4} \sum_i \frac 1{r_i} \g^{\mu\nu}_{(i)} \Box_{(4)} \d r_i 
\ = \  \cP^{\mu\nu;\rho\s}_{(4)}\ T_{\rho\s}\ 
\label{einstein-t2}
\end{align}
in agreement with \eq{einstein-equations-4D},
setting $F=0$ and taking into account \eq{t-trace-dim}.
Here $\cP$ is given explicitly by
\begin{align}
 \cP^{\mu\nu;\rho\s}_{(4)} &= 
 \L_0^{-4}\, e^{-2\s_4} \sum_i \frac 1{r_i^2}\,\g^{\mu\nu}_{(i)}\Big(\g^{\rho\s}_{(i)} - \frac 12\,\g^{\rho\s} \,\Big) .
\label{P-explicit-T2}
\end{align}
For $T^{\mu\nu} = 0$, we obtain indeed 2 propagating gravitational degrees of freedom
encoded in $\d r_4$ and  $\d r_5$.
The coupling to matter will be studied next.
To simplify the expressions, 
we will set $r_4 = r_5 = \frac{c}{2\pi} \equiv r_\cK$ from now on, so that the NC scale is defined 
appropriately as 
$\LNC^{-2} = 2\pi r_\cK\theta$ in the parametrization \eq{theta-6D}. 
Then we find from \eq{sigma4-explicit-torus}
\begin{align}
V_0\sqrt{|\theta^{ab}|}& %= V_0\theta^3 c 
 = \LNC^{-6}\, r_\cK^{-2} , \qquad
 \sqrt{|\g^{\mu\nu}_{(4-d)}|} = \LNC^{-8} , \qquad
  e^{\s_4}  = \LNC^{-2} \, r_\cK^2 
\label{dets-explicit}
\end{align}
so that the Newton constant is obtained as in \eq{Newton-const-4d}
\begin{align}
G_N &=  2\L_0^{-4} \, r_\cK^{-2} \ = \ 2 e^{\s_4} g_{\YM}^2\,r_\cK^{-2}
\ = \ 2 g_{\YM}^2\,\LNC^{-2} \ 
\label{G-newton-4D}
\end{align} 
up to factors of order 1.
Remarkably, the compactification radius $r_\cK$ drops out, and the 4-dimensional Planck scale is set by the 
noncommutativity scale and the gauge coupling constant.

\paragraph{Gravitational field of a point particle.}

Now consider a point mass $m$ on the above background, 
moving along  a time-like straight trajectory $v$,   %$v = v^i e_{(i)},\ i = 0,1,4,5$,
with a localized energy-momentum tensor  $T_{\mu\nu} \propto m v_\mu v_\nu$
and $v^2 = G^{\mu\nu}_{(4)}v_\mu v_\nu = -1$.
We can go to coordinates where the particle is at rest located at $\vec x = 0$.
To obtain the effective metric perturbation caused by $m$, we solve equation \eq{dr-eqn-t2}
\begin{align}
 \Box_{(4)} \d r_i &= \Delta_{(3)}  \d r_i =  m \,\L_0^{-4}\, \frac 1{r_\cK}\,
\big(v_{(i)}^2 + \frac 14\big)\,  \d^{(3)}(\vec x), \qquad
v_{(i)} = (e_{(i)},v)_G \ .
\label{dr-eqn-expl}
\end{align}
For simplicity we assume that $v_{(2)},\, v_{(3)} \approx 0$. Then the solution is
\begin{align}
 \frac{\d r_i(x)}{r_\cK} &= \ -  e^{\s_4}  g_{\YM}^2 \, \frac 1{r_\cK^2}\, \frac m{|\vec x|} 
 \ = \ -  \frac 12 G_N \, \frac m{|\vec x|}
\end{align}
This means that the radius of the torus  decreases in the presence of a mass,
and the classical Schwarzschild radius corresponds to  $\d r_i = - r_\cK$ i.e.  $r_i \approx 0$.
The resulting 4-dimensional metric perturbation is obtained as
\begin{align}
  \d  G^{\mu\nu}_{(4)} &= e^{-\s_4}\d \g^{\mu\nu}  - G^{\mu\nu}_{(4)} \d\s_4 
 = 2 e^{-\s_4}\frac 1{{r_\cK}}\sum_i\g^{\mu\nu}_{(i)}\d r_i - \frac 1{{r_\cK}} G^{\mu\nu}_{(4)} (\sum_i\d r_i) \nn\\
 \d  G_{\mu\nu}^{(4)} &= - 2 e^{-\s_4}\frac 1{{r_\cK}}\sum_i G_{\mu\mu'}^{(4)}G_{\nu\nu'}^{(4)}\g^{\mu'\nu'}_{(i)}\d r_i 
 + \frac 1{{r_\cK}} G_{\mu\nu}^{(4)}  (\sum_i\d r_i)
\end{align}
recalling that $\d  G_{\mu\nu} = - G_{\mu\mu'}G_{\nu\nu'}\d  G^{\mu\nu}$.
Therefore an observer at rest with respect to the source particle $m$, thus with velocity $v$, feels a 
static gravitational potential given by 
\begin{align}
 V(x) \ &= \ -v^\mu v^\nu \d G_{\mu\nu}^{(4)}\ =\  -2  m \, e^{\s_4}\,
\frac 1{r_\cK^2}(v_{(4)}^2 + v_{(5)}^2 -  v^2)\, \frac 1{|\vec x|} \nn\\
 &\approx  - G_N\, \frac {m}{|\vec x|}, \qquad G_N = 2  g_{\YM}^2 \, \LNC^{-2}
\end{align}
as long as $v_{(i)}^2 \ll 1$, using \eq{dets-explicit}.
This is indeed the attractive gravitational potential of a point mass, with
Newton constant \eq{G-newton-4D}.
Notice that the main contribution for this potential comes from the
trace contribution in \eq{P-explicit-T2}.

Comparing \eq{einstein-t2} with the Einstein equations,
we can consider the rhs as an effective modified energy-momentum tensor $\cP T$. 
In the above example, it implies that the point mass behaves like a 
particle with anisotropic pressure in general relativity. 
While the weak equivalence principle (universality of the metric)
essentially holds\footnote{There might be slight violations 
 due to the dilaton or the non-standard spin connection for fermions.}, 
the strong equivalence principle is clearly violated, because the 
tensor $\cP$  is not Lorentz invariant here. 
Spherical symmetry and isotropy might be restored either
in more sophisticated compactifications as discussed in the next section,
or possibly via the contributions from $F$. 
We can also verify the  weak energy condition for $\cP T$,
\be
 v'_\mu v'_\nu (\cP T)^{\mu\nu} \geq 0
\ee
for any time-like vector $v'$ on $M^4$. This is indeed satisfied,
\begin{align}
 v'_\mu v'_\nu (\cP T)^{\mu\nu} &= m\, \L_0^{-4}\,\frac 1{r_\cK^2}\, e^{-\s_4} \sum_i v'_\mu v'_\nu\g^{\mu\nu}_{(i)}
\Big(v_{(i)}^2 + \frac 12 \,\Big) \nn\\
 &= m\, \L_0^{-4}\,\frac 1{r_\cK^2}\,  \sum_i (v')_{(i)}^2  \Big(v_{(i)}^2 + \frac 12 \,\Big) \geq 0 \ .
\end{align}
However it should be kept in mind that the $F$ contributions are neglected here, 
which is not justified in the presence of $S_{\rm mix}$. Therefore the above 
treatment should not be taken too literally, but it serves to illustrate the mechanism. 
A more complete treatment will be given elsewhere.

\subsection{Discussion}

Let us discuss some aspects of the resulting gravity theory.
The basic question is if realistic (linearized) gravity can be recovered along these lines. 
In the simplest treatment above, $\cP$ is anisotropic,
and only certain components of $T^{\mu\nu}$ couple to gravity.
However, this might be fixed in various ways, such as a more sophisticated compactification,
or by taking the $F$ contributions and their mixing with the $\phi^A$ properly into account. 

It is  interesting to note that the 4-dimensional Planck scale
$\L_{\rm planck}^2 \sim G_N^{-1} \sim g_{\YM}^{-2} \LNC^{2}$  is indeed determined by the 
scale of noncommutativity, 
as may have been expected on naive grounds.  In particular,
the weakness of gravity arises  as a natural self-consistency condition for the semi-classical compactified
geometry. 
Notice also that no cosmological constant arises in \eq{einstein-equations-4D}.
This does not rule out however possible cosmological solutions with a similar behavior.
Indeed the mechanism also applies if $\cM^4 \subset \R^{10}$ has extrinsic curvature, which may
be interesting in the context of cosmology, as illustrated\footnote{These solutions 
assume a certain complexification of the Poisson structure which we
do not adopt here. However it seems plausible that similar
types of solutions exist also in the present setup.} in  \cite{Klammer:2009ku,Steinacker:2009mp}.
The role of quantum fluctuations will be discussed below.

\paragraph{Structural aspects.}

Some  structural remarks are in order. First,
one might worry that the lack of manifest diffeomorphism invariance of the matrix model 
leads to inconsistencies such as ghosts. This is not the case. The simplest way to see this is to
view the same model locally as $U(1)$ NC Yang-Mills theory on $\R^{2n}$.
Then there are massless propagating gauge and scalar fields
after performing the usual gauge fixing procedure, and consistency is manifest. 
From the geometric point of view, the point is that the metric fluctuations are 
automatically in harmonic gauge \eq{h-harmonic} in the matrix coordinates,
so that they are physical apart from the pure gauge contributions corresponding to
symplectomorphisms or
 would-be $U(1)$ gauge transformations, which are not part of the physical Hilbert space. 
 The significance of these geometrical modes was discussed in section \ref{sec:quadratic}.

% Let us discuss the geometrical degrees of freedom in more detail. There are $10-2=8$ physical degrees of freedom 
% on $\cM^{2n} \subset \R^{10}$ in the $U(1)$ sector of the IKKT model, taking into account gauge invariance.
% They split into $2n-2$ tangential and $10-2n$ transversal degrees of freedom. 
% We have seen that the transversal  moduli $\phi^A$ of $\cK$ are clearly gravitational modes.
% The remaining 2 transversal modes (in the case of $2n=6$) may be interpreted as sterile scalar fields, which may
% lead to non-standard gravitational effects such as gravity bags \cite{Steinacker:2009mp}.
% This seems at odds with  previous proposals
% \cite{Rivelles:2002ez,Yang:2006hj,Steinacker:2010rh} to interpret the two propagating $A_\mu$ on $M^4$ 
% in terms of gravitons. 
% However, two of these tangential degrees of freedom may be absorbed in the two scalar fields 
% $e^{-\s}$ and $\eta \sim G^{\mu\nu}g_{\mu\nu}$ corresponding to dilaton and axion, and  
% others may be frozen by the flux condition for $\cK$. 
% In the presence of matter, these tangential perturbations $F$ may contribute to the
%  6 geometrical degrees of freedom required for the most general 4-dimensional effective metric. 

Note also that
the vacua under consideration break the global $SO(6)$ symmetry of the model, and can be viewed
in terms of time-dependent VEV's of scalar fields
 from a 4-dimensional NC field theory perspective.
Accordingly, some of the massless would-be $U(1)$ modes can be viewed as Goldstone bosons resulting from the
breaking of the global $SO(6)$ (or even $SO(9,1)$) symmetry of the model by the background, although
this analogy goes somewhat beyond\footnote{cf. also \cite{Nicolis:2011pv} for a somewhat related recent discussion.} 
the standard field-theoretical setting due to the presence of $\theta^{\mu i}$.

\paragraph{Relation with string theory.}

From a string theory point of view,
it is remarkable that the effective gravity is  4-dimensional, even though the brane
$\cM^4 \times \cK \subset \R^{10}$ is embedded in a higher-dimensional non-compact target space.
This is in contrast to the conventional picture where gravity originates from 
closed strings which propagate in 10 dimensions, leading to a 10-dimensional Newton law if embedded in $\R^{10}$.
The crucial point here is that the effective brane gravity is governed by the open string metric
which encodes a non-degenerate $B$-field, realizing split noncommutativity $\theta^{\mu i} \neq 0$ and
large extrinsic\footnote{In particular,  an abstract DBI-type action for the brane 
would not reproduce the above mechanism unless the induced ''closed string`` metric on the 
brane is properly realized as an embedding metric.} curvature of $\cK\subset\R^{10}$. Then the compactification moduli
couple appropriately to 4-dimensional matter and mediate brane gravity. In contrast, 
the bulk gravity arises in a holographic manner. 
This origin for a 4-dimensional behavior is very different from e.g. the DGP mechanism \cite{Dvali:2000hr},
which is based on a combination of brane and bulk physics with Einstein-Hilbert term but without a $B$ field.

The fact that 4-dimensional gravity can arise on branes in a non-compact bulk is very interesting. 
It means that there is no need to consider the vast landscape of 6-dimensional string compactifications
with its inherent lack of predictivity. Rather, there is a mini-landscape of at most 4-dimensional
compactifications $\cM^{2n} = \cM^4 \times \cK \subset \R^{10}$, which is not only 
much smaller but also governed by a meaningful selection mechanism given by the matrix model.
In principle one can even put the model on a computer, which has recently 
provided interesting evidence in favor of effectively 4-dimensional 
vacuum geometries \cite{Kim:2011cr}.

Some remarks on the claimed UV finiteness are in order.  This claim is based on two grounds: 1) the compactified 
brane backgrounds $M^4 \times \cK \subset \R^{10}$ behave in the UV as 4-dimensional noncommutative spaces \cite{Steinacker:2011wb},
and 2) the IKKT model is equivalent to $\cN=4$ NC SYM on a 4-dimensional background, and thus free of pathological UV/IR mixing and 
perturbatively finite, cf. \cite{Jack:2001cr}.
Point 1) is very intuitive, since compact NC spaces can carry only finitely many degrees of freedom,
and shown explicitly in \cite{Steinacker:2011wb}.
Point 2)  needs to be confirmed more rigorously, but is very reasonable. 
Note that  we consider the matrix model as fundamental and independent from string theory, hence there are no other degrees of freedom
apart from the ones captured by NC gauge theory. This may deviate from string theory, which contains also an infinite 
tower of closed string modes whose relation to the matrix model is unclear; the relation with NC field theory is established
only in a suitable $\a'\to 0$ limit \cite{Seiberg:1999vs}. I fact there is no claim for perturbative finiteness in the 
matrix model for genuinely higher-dimensional backgrounds such as $\R^6_\theta$ or $\R^{10}_\theta$,
as already pointed out in \cite{Ishibashi:1996xs}. 
Therefore the present claim to obtain a (perturbatively) UV finite theory including gravity is based on the 
assumption of  compactified\footnote{It is important to note that due to their non-commutative nature,
the branes of type $\R^4\times K$ as considered here become effectively 4-dimensional in the UV, 
as shown explicitly in \cite{Steinacker:2011wb}. This is a consequence of the 
uncertainty relations along with the compactness of $\cK$.} 4-dimensional brane solutions, 
and is independent of the UV finiteness of string theory.
This may even be viewed as an argument in favor of 4-dimensional branes in the IKKT model.

Finally, while the matrix model is considered as fundamental here,
the same mechanism should also arise in the context of  IIB string theory
in a suitable decoupling limit,
for a brane $\cM^4 \times \cK \subset \cM^{10}$ with  non-degenerate $B$-field corresponding to split noncommutativity
embedded in a non-compact $\cM^{10}$.
Then the bare matrix model action would be replaced by the Dirac-Born-Infeld action.
It would be very interesting to understand the resulting modifications, which 
should capture quantum corrections within the matrix model approach.

\paragraph{Vacuum energy and the cosmological constant problem.}

It is remarkable that the geometric equations \eq{einstein-equations-4D} resp. \eq{einstein-t2}
do not involve any cosmological constant. 
However, the energy-momentum tensor might of course contain a component $T_{\mu\nu} \propto G_{\mu\nu}$
induced by quantum fluctuations, which typically happens upon quantization. 
Since the model is supersymmetric, only modes below the SUSY breaking scale 
$\L_{\rm SUSY}$ contribute, so that $T_{\mu\nu}^{(\rm vac)} \sim \L_{\rm SUSY}^4 G_{\mu\nu}$. 
This would modify equation \eq{einstein-equations-4D} with a cosmological constant term similarly as in GR,
and it appears that the usual cosmological constant problem would arise. However, this conclusion
is premature. The structure of the compactified vacuum geometry
$\cM^{2n} = \cM^4 \times \cK$ must be determined by taking into account all contributions to the effective action,
including such quantum effects. While this is beyond the scope of the present paper,
we can give a simple argument in favor of solutions with flat 4-dimensional
geometry $\cM^4$ and constant compactification, even in the presence of vacuum energy. To this end,
note that the semi-classical action \eq{action-semiclass} $S\sim\L_0^4\int \sqrt{G}\, G^{ab}g_{ab}$
has a similar structure as the induced vacuum action 
$S_{\rm vac} \sim\L_{\rm SUSY}^4\int \sqrt{G}$. 
It is then easy to see that the equations of motion obtained from the combined action
$S = \int \sqrt{G}\,(\L_0^4 G^{ab}g_{ab} +\L_{\rm SUSY}^4)$ take the form
\be
\Box_{\tilde G}\, x^a = 0, \qquad \tilde G^{ab} = G^{ab} + \a\frac{\L_{\rm SUSY}^4}{\L_{0}^4} g^{ab} .
\ee
This has the same structure as the bare e.o.m. used throughout this paper, with a small modification of the 
effective metric suppressed by $\frac{\L_{\rm SUSY}^4}{\L_{0}^4}$. 
It is therefore very plausible that the main results of this paper apply also upon taking into account 
vacuum energy, and at least nearly-flat vacuum geometries should exist even in the presence of vacuum energy.
In a similar vein, the full Dyson-Schwinger equations of the quantized matrix model take the form of the
bare matrix model equations $\langle\Box X^A\rangle =0$ (dropping fermions).  
These arguments strongly suggest that the cosmological constant problem
may be less serious or even resolved in the present approach. 
However this require a careful study of the model at the quantum
level, which is beyond the present paper.

\section{More general compactifications}
\label{sec:generalizations}

\subsection{Effective 4-dimensional metric and averaging}

One problem of the above background is that the radial moduli correspond to specific
metric degrees of freedom, which couple to the energy-momentum tensor via $\cP$ in an anistropic way. 
This problem may be alleviated for more general compactifications, where
the metric components $\g^{\mu\nu}_{(i)}$ are not constant along $\cK$ (or $\cM$) but 
rotate along $\cK$. Then the tangential moduli on $\cK$ may also play a significant role for gravity.

To understand the effective 4-dimensional metric
on more general compactified backgrounds $\cM^4 \times \cK$, we can decompose the fields in harmonics 
 i.e.  Kaluza-Klein modes on $\cK$, and restrict ourselves to the lowest KK mode.  
For example, consider a scalar field $\varphi = \varphi(x^\mu)$ 
which is constant along the compact space $\cK$. 
Then the action in the M.M. reduces to
\begin{align}
S[\varphi] &=  \int d^{2n} x \sqrt{| G|} \, G^{ab} \del_a\varphi \del_b \varphi  
= \int d^{4}x \Big(\int_K d\zeta\,\sqrt{\det\th^{-1}_{ab}} \, \g^{\mu\nu}\Big) \del_\mu\varphi \del_\nu \varphi .
\label{S-phi-higherdim-red}
\end{align}
Assuming that $\det{\th^{-1}_{ab}}$ is constant along $\cK$, we 
 define a reference volume $V_0$ of $\cK$ via\footnote{Assuming a product structure $\cM^4\times \cK$, one can
integrate the symplectic measure over $\cK$ and obtain a volume form on $\cM^4$.
We assume that this can be done via e.g. some fixed canonical invariant measure on a 
homogeneous space indicated by $\int_K d\zeta\,$.}
\begin{align}
\int d^{2n} x\sqrt{\det\th^{-1}_{ab}} = \int d^{4} x \int_K d\zeta \sqrt{\det\th^{-1}_{ab}}
 =:\int d^{4} x\, V_0\sqrt{\det\th^{-1}_{ab}}
\end{align}
as anticipated previously.
We do not require that $\g^{\mu\nu} =  \theta^{\mu a}\theta^{\nu b} g_{ab}$ is 
constant along $\cK$.  
Then the effective 4-dimensional metric is determined by the reduced $2n$-- dimensional conformal metric
averaged  over $\cK$.
The appropriate scale factor of the effective 4-dimensional metric is 
determined as in \eq{eff-metric} such that
\begin{align}
S[\varphi] &= \int d^{4}x \sqrt{| G_{(4)}|} \, G_{(4)}^{\mu\nu} \del_\mu\varphi \del_\nu \varphi \ ,
\end{align}
which leads to
\begin{align}
  G_{(4)}^{\mu\nu} &=  e^{-\s_{4}} \, \bar \g^{\mu\nu}, \qquad \bar \g^{\mu\nu} = \int_\cK \frac{d\zeta}{V_0}\, \g^{\mu\nu} ,  \qquad
e^{-\s_{4}} =\frac{V_0\,\sqrt{|{\th^{-1}_{ab}|}}}{\sqrt{|{G^{(4)}_{\mu\nu}|}}}\, .
\label{g-4-normalization}
\end{align}
Therefore $G_{(4)}^{\mu\nu}$ is the effective metric which governs the 4-dimensional 
physics of the lowest KK modes.
It follows as in \eq{dGs-torus} that
\begin{align}
 \d G_{(4)}^{\mu\nu} &= e^{-\s_4} \d \bar\g^{\mu\nu} - G_{(4)}^{\mu\nu} \d \s_4, \qquad
\d \s_4 = \frac 12 \bar\g_{\mu\nu}\d \bar\g^{\mu\nu}\, .
\end{align}
The coupling to matter
can be written  for the lowest Kaluza-Klein modes either
in $2n$ or in $4$ dimensional form,
\be
\begin{array}{rll}
\d S_{\rm matter} &= \int d^{2n}x \sqrt{G_{ab}} \, \d G^{ab} T_{ab} \
 &= \ \int d^{4}x \sqrt{G_{(4)}} \, \d G_{(4)}^{\mu\nu} T^{(4)}_{\mu\nu}  \nn\\
&=  \int d^{2n}x \sqrt{G_{ab}} \, (e^{-\s} \d \g^{ab} - G^{ab} \d \s) T_{ab} \
 &= \ \int d^{4}x \sqrt{G_{(4)}} \,(e^{-\s_4} \d \bar\g^{\mu\nu} - G_{(4)}^{\mu\nu} \d \s_4)  T^{(4)}_{\mu\nu} .
\end{array}
\label{coupl-matter-4d-gen}
\ee
At first sight, there appears to be a mismatch for the trace contribution,
since $\d \s = \frac 1{2(n-1)} \g_{ab}\d \g^{ab}$ while $\d \s_4 = \frac 1{2} \bar\g_{\mu\nu}\d \bar\g^{\mu\nu}$.
This is resolved by noting that 
\be
\int d^{2n}x \sqrt{G_{ab}} \,T_{ab} G^{ab} =  \int d^{4}x \sqrt{G_{(4)}} \,(n-1) T^{(4)}_{\mu\nu} G^{\mu\nu}_{(4)} 
\label{t-trace-dim}
\ee
provided $\varphi$ and $\d \bar\g^{\mu\nu}$ are constant along $\cK$;
this is easily checked e.g. for scalar fields, 
\begin{align}
T_{ab} =  \del_a\varphi\del_b\varphi - \frac 12 G_{ab} (G^{cd}\del_a\varphi\del_b\varphi),
 \quad  T_{\mu\nu}^{(4)} =  \del_\mu\varphi\del_\nu\varphi - \frac 12 G_{ab}^{(4)} (G^{cd}_{(4)}\del_a\varphi\del_b\varphi) 
\nn
\end{align}
while $T=0$ for gauge fields. 
The last form of \eq{g-4-normalization} implies as in \eq{einstein-tensor-fluct}
that the corresponding 4-dimensional Einstein tensor can be written as
\begin{align}
\cG_{(4)}^{\mu\nu} = \frac 12 e^{-\s_4} \bar\Box_{(4)} \d\g^{\mu\nu}  \ + \cO(F) ,
\end{align}
using again a suitably adapted flat background, 
and using the harmonic gauge condition \eq{h-harmonic-4d}.
For constant curvature compactifications as above, this reduces to
\begin{align}
\cG_{(4)}^{\mu\nu} &= e^{\s-2\s_4} \sum_i \frac 1{r_i}\int_K \frac{d\zeta}{V_0}\, \bar \Box(\g^{\mu\nu}_{(i)} \d r_i ) \ +  \ \cO(F) \nn\\
 &=  e^{\s-2\s_4} \sum_i \frac 1{r_i}  \int_K \frac{d\zeta}{V_0}\,\g^{\mu\nu}_{(i)} \bar\Box \d r_i   \ +  \ \cO(F) \nn\\
 &= \cP_{(4)}^{\mu\nu;\rho\eta}\ T_{\rho\eta}\ +  \ \cO(F)
\label{einstein-equations-moduli-average}
\end{align}
assuming  $\nabla\gamma_{(i)}=0$ or $\del|_p \d r_i=0$, and\footnote{These manipulations are somewhat sketchy, 
and should be refined elsewhere.} the $2n$-dimensional equation of motion  \eq{eom-dr} for the moduli.
The $\cO(F)$ term will be discussed in the next section.
Here
\begin{align}
 \cP_{(4)}^{\mu\nu;\rho\eta}&= 
 \L_0^{-4}\, e^{-2\s_4} \sum_i \frac 1{r_i^2}\, \int_K \frac{d\zeta}{V_0}\,\g^{\mu\nu}_{(i)}
\Big(\g^{\rho\eta}_{(i)} - \frac{\dim \cK_i}{2(n-1)}\, \g^{\rho\eta} \,\Big) .
\label{einstein-equations-4D-general}
\end{align}
In particular, we recover  \eq{einstein-equations-4D}.
The important point is that $\cP$ is now averaged over $\cK$, and the partial metrics $\g^{\mu\nu}_{(i)}$
 probe different components of the energy-momentum tensor $T_{\mu\nu}$.
This or a similar averaging
might allow to make the present mechanism of brane gravity realistic.

Notice that  we used above the 4-dimensional harmonic gauge condition 
\be
\partial^\nu  h^{(4)}_{\mu\nu} - \frac 12 \partial_\mu h^{(4)} =0 .
\label{h-harmonic-4d}
\ee 
This follows again from the 4-dimensional equations of motion. Indeed, the variation of the action for 
any scalar field $\varphi$ which is constant along $\cK$ %(such as $x^\mu$)
 can be written as 
\begin{align}
  \d_{(4)} S[\phi] = 2\int d^{4}x  \,\d_{(4)}\varphi \partial_\nu (\sqrt{| G_{(4)}|}\,  G_{(4)}^{\nu\mu}\del_\mu \varphi)
= 2\int d^{4}x \sqrt{| G_{(4)}|} \,\d_{(4)}\varphi \Box_{(4)} \varphi
\end{align}
Comparing with \eq{S-phi-higherdim-red}, it follows that 
the Laplace operator %$\Box \varphi = \star d \star d \varphi$ 
for the lowest KK modes reduces to that of $G_{(4)}$:
\begin{align}
\sqrt{| G_{(4)}|}\,\Box^{(4)}_{G} \varphi 
\ =\  V_0\sqrt{| G_{(2n)}|}\,\Box^{(2n)}_{G} \varphi .  
\label{laplace-4-2n}
\end{align}
In particular, the matrix coordinates $x^\mu$ 
are harmonic also  w.r.t. $G_{(4)}$ in vacuum,
\begin{align}
 \Gamma^\mu_{(4)}= \Box^{(4)} \, x^\mu 
 = V_0\frac{| G_{(2n)}|^{1/2}}{|G_{(4)}|^{1/2}}\,\Box_{G} x^\mu 
 = e^{\s-\s_4}\, \Gamma^\mu_{(2n)} = 0 .
\label{harmon-reduced}
\end{align}
Therefore the harmonic gauge condition \eq{h-harmonic-4d} holds. 

We conclude that the massless moduli lead to (nearly) Ricci-flat deformations of the 
effective 4-dimensional metric, not only for toroidal compactifications but also under somewhat weaker assumptions
on the compactification. The coupling of gravity to matter will quite generically lead to Newtonian gravity, and
 a (near-) Lorentz invariant 4-dimensional effective $\cP$ might be recovered 
upon averaging over $\cK$, possibly leading to a  viable gravity close to GR.
A similar averaging may arise 
in the presence of multiple branes with intersecting compactifications $\cK$, where
each brane will contribute gravitational modes, which may couple to
different components of the energy-momentum tensor. Such scenarios are very appealing also from the particle physics
point of view, and are naturally realized in matrix models \cite{Chatzistavrakidis:2011gs}.
On the other hand, compactifications with $\nabla\theta^{ab} \neq 0$ may also be interesting,
as  discussed next.

\subsection{Non-constant $\theta^{ab}$}
\label{sec:non-constant}

The toroidal compactification considered above are special because the Poisson tensor is covariantly constant.
This implies that the metric fluctuations due to $F$ couple only to derivatives of the e-m tensor. 
We briefly discuss the effect of more general compactifications with $\nabla\theta^{ab}\neq 0$ on $F$.
The equations of motion \eq{eom-A-T} now imply
\begin{align}
\bar\Box F_{ab}\ &=\ 2\L_0^{-4}\,e^{-\sigma}\,
 \Big(T_{ef} \,\bar G^{fc}(\bar\nabla_b\bar\nabla_c\bar\theta^{ed} \bar G_{da} - \bar\nabla_a\bar\nabla_c\bar\theta^{ed}\bar G_{db}) +\cO(\nabla T) \Big)  .
\label{box-F-T-rotating}
\end{align}
Unlike for toroidal compactifications,
the tangential fluctuations now  couple to the e-m tensor and not just its derivatives, and 
 may play a similar role as $\phi^A$ for gravity. 
Computing the 4-dimensional Einstein tensor for the averaged  metric 
\begin{align}
\cG^{ab}  
 &=  e^{-\s}\sum_i\frac 1{r_i}\bar\gamma_{(i)}^{ab} \bar\Box \d r_i
 - \frac 12 \bar G^{bd} \bar\Box (\bar\theta^{ac} F_{cd}) - \frac 12 \bar G^{ad} \bar\Box (\bar\theta^{bc} F_{cd})
 + \frac 14 \bar G^{ab} \bar\Box (\bar\theta^{cd}F_{cd}) \  \nn\\
 &  \ \ + \cO(\nabla\nabla T) \ + \cO(\d^2) 
\label{einstein-equations-mod-gen}
\end{align}
will lead to an Einstein-type equation
\begin{align}
\cG^{\mu\nu}  &= (\cP^{\mu\nu;\rho\eta}_{\phi} + \cP^{\mu\nu;\rho\eta}_F)\ T_{\rho\eta} \  + \cO(\nabla\nabla T) \ + \cO(\d^2) .
\label{P-mod-gen}
\end{align}
We assume that the $\bar\nabla\bar\theta\bar\nabla F$ terms vanish upon averaging over $\cK$,
which needs to be verified in the specific compactifications.
Here $\cP^{\mu\nu;\rho\eta}_\phi$ is as before, and 
\be
 \cP^{\mu\nu;\rho\eta}_F = \L_0^{-4}e^{-\s}\cO(\bar\theta\nabla\nabla\bar\theta) = \cO(G_N) 
\ee 
describes the contribution of the would-be $U(1)$ gauge fields due to 
the first term in \eq{box-F-T-rotating}.  Then
both transversal and tangential perturbations contribute with the same coupling 
strength $G_N$. 
Hence for such compactifications $\cK$ the tangential modes  play a similar role as the 
transversal modes, and a 4-dimensional  $\cP$ 
providing an appropriate coupling to the full 4-dimensional e-m tensor
might be recovered upon averaging over $\cK$.
Thus after the dust has settled, the present 
scenario of emergent gravity from the IKKT model might provide a viable description of 
gravity and its quantization.

\section{Conclusion}

In this paper, a new mechanism is exhibited which leads to effective 4-dimensional gravity on compactified 
brane solutions $\cM^4 \times \cK \subset \R^{9,1}$
of the IKKT matrix model. Gravitational modes are encoded in the 
moduli of the compactification, which are transmitted to the non-compact space via 
the Poisson tensor. The required type of Poisson structure (dubbed split noncommutativity)
arises naturally in the relevant solutions, so that the mechanism is robust and natural.
No Einstein-Hilbert action is required, only the basic matrix model action is used
which implies a certain type of harmonic embedding. It turns out that 
the Einstein tensor is indeed sourced by the energy-momentum tensor. Although
vacuum geometries are not necessarily Ricci-flat, we argue 
that Ricci-flat vacuum geometries do arise naturally at least in the linearized regime.
The gravitational coupling depends on the specific compactification  $\cK$.
For the simplest case of $\R^{3,1} \times T^2$, the  coupling is anisotropic a priori. 
We argue that more sophisticated compactifications and/or a more complete treatment of
the Poisson structure should lead to a
physically viable effective gravity theory in 4 dimensions.
Although the model is closely related to $\cN=4$ SYM,
the mechanism does not arise in the  commutative setting, because the 
Poisson tensor $\theta^{i\mu}$ provides the essential link between the moduli of $\cK$ and the 
non-compact metric. 

There are several reasons why this non-standard origin for gravity is interesting. 
First of all, it promises to give a perturbatively finite quantum theory of 
gravity. The reason is that the compactified backgrounds under consideration 
behaves in the UV like 4-dimensional spaces, due to their intrinsic noncommutative nature.
This should imply perturbative finiteness, because the model can be viewed
alternatively as $\cN=4$ NC gauge theory on a  4-dimensional NC space. 
%Note that no reference to an infinite tower of stringy states is made here. 
Furthermore, the model offers reasonable hope to resolve the cosmological constant problem, 
and we argued that vacuum energy induced by quantum mechanical zero-point fluctuations 
should be consistent with flat 4-dimensional geometries. 
Finally, this approach avoids the landscape problem in string theory, 
because it does not require 6-dimensional compactifications but only 2- or 4-dimensional compactification.
Of course all these claims are bold, and require careful scrutiny and better justification.
Nevertheless they appear to be reasonable, and certainly justify to study this matrix-model 
approach in detail.

This paper clearly leaves many open questions and loose ends. 
The main point is to demonstrate the mechanism, and to provide hints for further explorations.
There are many obvious directions for follow-up work, in particular
more sophisticated compactifications, non-trivial embeddings of the 
4-dimensional space, and a more complete treatment of the
 tangential modes. The same mechanism should also be studied form other points of view,
such as the BFSS model, or from a more stringy perspective. 
%This should also make clear that the present mechanism evades the Weinberg-Witten theorem.
It remains to be seen if a fully realistic quantum theory of gravity and 
other fundamental interactions will emerge form this approach.

\paragraph{Acknowledgments.}

I would like to thank in particular A. Polychronakos and the theoretical high-energy physics group of the 
City University of New York for hospitality, very useful discussions and support. 
This work  was supported in part by the Austrian Science Fund (FWF) under the contract
P21610-N16, and in part by a CCNY/Lehman CUNY collaborative grant.

\startappendix

\Appendix{Extrinsic curvature}
\label{sec:ext-curv}
%%%%%%%%%%%%%%%%%%%%%%%%%%%%%%%%%%%%%%%%

Consider the objects
\begin{align}
K_{ij}^A &=  \ \nabla_{i} \del_{j} x^A, \qquad K_{ij;kl} \ = \ K_{ij}^A  K_{kl}^B\, \eta_{AB} 
\end{align}
on $\cM \subset \R^m$. Viewing the Cartesian embedding functions $x^A$  as scalar fields on $\cM$,
we can consider $K_{ij}^A$ as a rank 2 tensor field for each $A$, and 
$K_{ij;kl}$ as a tensor field. 
If $\nabla \equiv \nabla_G = \nabla_g$, then the Gauss-Codazzi theorem states that 
\begin{align}
 K_{ij;kl} - K_{jk;il}  = R_{ikjl} .
\label{Gauss-codazzi}
\end{align}
Let us compute these objects for the sphere $S^{m} \subset \R^{m+1}$.
Using the $SO(m+1)$ symmetry,
we can use normal embedding coordinates $x^i,\ i = 1,...,m\ $ for
\begin{align}
x^A = \begin{pmatrix}
 x^i, \qquad i = 1,...,m \\
 x^{m+1} = \sqrt{r^2-\sum (x^i)^2}
\end{pmatrix}
\end{align}
at the north pole $p = (0,.,0,r)$. Then
\begin{align}
 \del_i\del_j x^k &= 0, \quad k = 1,...,m, \qquad 
  \del_i\del_j x^{m+1} 
= - \(\frac{(x^{m+1})^2\d_{ij} + x^i x^j}{(x^{m+1})^3}\) ,
\end{align}
so that
\begin{align} 
  \nabla_i\del_j x^A|_p &= P_N  \del_i\del_j  x^A|_p
 =  \begin{pmatrix}  0 \\ - \frac{\d_{ij}}{R} \end{pmatrix}
\end{align}
which means that
\begin{align}
K_{ij}^A &= - \frac{g_{ij}}{r^2}\, x^A ,  \qquad
K_{ij;kl} = \frac 1{r^2}\, g_{ij}\, g_{kl} .
\end{align}
In particular for a torus $T^m = \times_a\ S^1_{(a)} \subset \R^{2m}$, we obtain
\begin{align}
K_{ij;kl} &= \sum_a\frac 1{r_a^2} g_{ij}^{(a)} g_{kl}^{(a)} 
\end{align}
where $r_a$ are the radii of the cycles of $T^m$.
One can then verify via the  Gauss-Codazzi theorem that the intrinsic geometry is flat.

\Appendix{Matrix energy-momentum tensor}
\label{sec:e-m-tensor}

We recall that the translations
$\d X^A = c^A\one$  are symmetries of the matrix model.
Adapting a standard trick from field theory, one can derive a corresponding conservation law by 
considering the following
non-constant infinitesimal transformation\footnote{it is not hard to see that this preserves the 
measure of the matrix path integral.} 
\be
\d X^A = X^B[X^A,\varepsilon_{B}] + [X^A,\varepsilon_{B}]X^B
\label{transl-diffeo}
\ee
where $\varepsilon_B$ is an arbitrary matrix.
As elaborated in  \cite{Steinacker:2008ri}, this leads to
\begin{align}
 \d S_{\rm YM} 
&= - Tr  \varepsilon_B [X_A, \cT^{AB}] 
= Tr  [X_A,\varepsilon_B] \cT^{AB}  
\label{T-cons}
\end{align}
for an arbitrary matrix (!) $\varepsilon_B$, so that 
\be
[X_B,\cT^{AB}]  =0
\label{e-m-cons}
\ee
 where $\cT^{AB}$ is the ''matrix`` energy-momentum tensor. Is bosonic contribution is given explicitly by
\be
 \cT^{AB} = \frac 12 ([X^A,X^C][X^B,X_{C}]  + (A \leftrightarrow B))  - \frac 14 \eta^{AB} [X^C,X^D][X_C,X_D] .
\label{e-m-tens}
\ee
Its $U(1)$-valued component $\cT^{AB} = \cT_{\rm geom}^{AB} + \cT_{\rm nonabel}^{AB}$ consists of a geometrical term 
\begin{align}
\cT_{\rm geom}^{AB} &= \del_a X^A \del_b X^B  \theta^{aa'}\theta^{bb'}\ T_{ab}^{\rm geom}, \qquad 
 T_{ab}^{\rm geom} = - g_{ab}  + \frac 14 G_{ab} (G^{cd}g_{cd})
\label{T-geom-def}
\end{align}
 plus the energy-momentum tensor
of the nonabelian gauge and scalar fields.
It is easy to see from \eq{e-m-tens} and \eq{box-def} that
this can be rewritten as
\begin{align}
 \{X_B,\cT_{\rm geom}^{AB}\} &= \{X^A,X_B\} \Box X^{B}
 =  e^\s  \theta^{ac}\del_a x^A g_{bc} \Box x^b 
 =  e^{2\s} G^{ab} G^{de}\nabla_e^{(g)}\theta^{-1}_{db} \, \del_a x^A 
\end{align}
which defines a vector field on $\cM$, using  the identity
\begin{align}
\Box_G x^b &= -\G^b =  \frac 1{\sqrt{G}}\nabla^{(g)}_a(\sqrt{G}G^{ab})
=  e^{-\s} \theta^{ac}g_{dc}\nabla_a^{(g)}\theta^{bd} 
% &=  e^{-\s} \theta^{dd'}\theta^{ac}g_{dc}\theta^{bb'}\nabla_a^{(g)}\theta^{-1}_{b'd'}
=  - G^{d'a}\nabla_a^{(g)}\theta^{-1}_{b'd'} \, \theta^{bb'} 
\end{align}
in the last step.
On the other hand, in any local coordinates we can write
\begin{align}
\{X_B,\cT^{AB}\} &= \theta^{cd}\del_c X^B \del_d \cT^{AD}  \eta_{BD} \nn\\
 &=  \theta^{cd}\del_c X^B \del_d (\del_a X^A \del_b X^D \theta^{aa'}\theta^{bb'}  T_{a'b'} )  \eta_{BD} \nn\\
&=  \theta^{cd} \del_d (\del_a X^A\del_c X^B \del_b X^D \theta^{aa'}\theta^{bb'}  T_{a'b'} )  \eta_{BD} \nn\\
&=  \del_d ( g_{cb} \theta^{bb'} T_{a'b'} \theta^{aa'}\del_a X^A)  \nn\\
&=  \sqrt{\theta} \del_d (\sqrt{\theta^{-1}} \theta^{cd} g_{cb} \theta^{bb'} T_{a'b'} \theta^{aa'}\del_a X^A)  \nn\\
&=  \frac{e^\s}{\sqrt{G}} \del_d (\sqrt{G} G^{db'}  T_{a'b'} \theta^{aa'}\del_a X^A) 
\label{cons-law-explicit}
\end{align}
using the identity \eq{theta-id}.  
In NEC or equivalently $\nabla^{(g)}$, the double derivative term $\del_d\del_a X^A$ is in the normal bundle, 
so that we obtain the  identity
\begin{align}
  \nabla^{(g)}_d (\sqrt{G} G^{db}  T_{a'b}^{\rm geom} \theta^{aa'}) 
 = e^{\s} \sqrt{G}\, G^{ab} G^{da}\nabla_a^{(g)}\theta^{-1}_{db} .
\label{geom-id-cons}
\end{align}
Finally,
to see the relation with the usual e-m tensor, note that the nonabelian components contribute via
\begin{align}
 [X^a,X^b] = \theta^{aa'}\theta^{bb'} (\theta^{-1}_{a'b'} + \cF_{a'b'})
\end{align}
and we recover the standard form of the e-m tensor for nonabelian gauge fields
\begin{align}
\cT_{\rm nonabel}^{ab} &= e^\s \theta^{a a'}\theta^{b b'} (\cF_{a' c}  G^{cc'} \cF_{c'b'} -\frac 14  G_{a'b'} (\cF\cF))
 \ = \ e^\s \theta^{a a'}\theta^{b b'} T_{a'b'}^{\rm nonabel}\ .
\end{align}

\Appendix{Harmonic correction due to matter.}
\label{app:harmonic}

In the presence of  matter, the equations of motion are modified so that the 
matrix coordinates are no longer harmonic, $\Box x^a \neq 0$. However, we will show  that the 
deviation from harmonicity is small and negligible compared with the energy-momentum tensor source 
for the Ricci tensor, justifying the above derivation of the geometric equations of motion.
To see this,
we recall the general expression \eq{lin-Ricci-general} for the linearized Ricci tensor  in terms of $h_{ab}$, 
which can be written as
\begin{align}
 \d R^{ab} &=\frac 12 \nabla^a(\nabla_d h^{bd} -\frac 12\del^b h) +  \frac 12 \nabla^b(\nabla_d h^{ad} -\frac 12\del^a h)
- \frac 12 \Box h^{ab} \quad \nn\\
&= -\frac 12 \nabla^a \Box x^b -\frac 12 \nabla^b \Box x^a - \frac 12 \Box h^{ab} \quad
\label{lin-Ricci-general-2}
\end{align}
since
\begin{align}
\Gamma^a = -\Box x^a \sim \nabla_d h^{ad} -\frac 12\del^a h 
\end{align}
(note that $h^{ab} = - \d  G^{ab}$). 
% Now the equations of motion of the matrix model are obtained by variations
% $X^A \to X^A + \d X^A$, 
% \begin{align}
% \d S = \Tr \d X_A(\Box X^A \ + \frac{\d S_\Psi}{\d X^A} ) 
% = \Tr \d X_A(\Box X^A \ - \frac 12 [\obar\Psi,\gamma^A\Psi]) ,
% \end{align}
% which gives the following on-shell relation
% \begin{align} 
%  \Box X^A \ &= \frac{\d S}{\d X^A} + \frac 12[\obar\Psi\gamma^A,\Psi]  \ . 
% \end{align}
% The point is that the last term is a double derivative term due to noncommutativity, and negligible compared with $T^{ab}$. 
% The same applies to contributions due to nonabelian fields.
Now note that the $U(1)$ component of the 
conservation law \eq{e-m-cons} casts the tangential equations of motion 
in the presence of matter in the following useful form
\begin{align}
e^\s \G^d = - e^\s\Box x^d = g^{bd}\theta^{-1}_{bc} [X_B,\cT^{cB}_{\rm nonabel}] \ ,
\end{align}
and similarly for the fermionic matter contributions. 
Plugging this into \eq{lin-Ricci-general-2}, 
the terms $\nabla^a \Box x^b$ contribute second derivatives of the 
energy-momentum tensor resp. of $\theta^{ab}$, which is much smaller than the
matter contributions to $\Box h^{ab}$  which led to
\eq{einstein-equations-lin}, thus finally justifying its derivation in the 
presence of matter (at least if $\theta^{ab}$ is constant).
Notice that these derivative contributions are of the same magnitude as the contributions of the 
tangential perturbations $F$ in \eq{einstein-equations-lin}, so that the
derivative corrections to the Einstein equations in the presence of matter
require a more careful investigation.

%%%%%%%%%%%%%%%

\end{document}